\documentclass[jcp,twocolumn]{revtex4}
\usepackage{amsmath}
\usepackage{amsfonts}
\usepackage{float}
\usepackage{graphicx}
\usepackage{marginnote}

\newcommand{\dd}{{\ensuremath{\mathbf{d}}}}
\newcommand{\dtt}{{\ensuremath{d_{33}}}}

\newcommand{\firstDeriv}[2]{{\ensuremath{\frac{\partial #1}{\partial #2}}}}

\newcommand{\secondDerivMix}[3]{{\ensuremath{\frac{\partial^2 #1}{\partial #2 \partial #3}}}}

\newcommand{\uu}{{\ensuremath{\bf{u}}}}
\newcommand{\ff}{{\ensuremath{\bf{f}}}}
\newcommand{\vv}{{\ensuremath{\bf{v}}}}
\newcommand{\eq}[1]{{Eq. (\ref{#1})}}
\newcommand{\fig}[1]{{Fig. \ref{#1}}}

\newcommand{\bld}[1]{\ensuremath{{\mathbf #1}}}

\begin{document}

\title{Developing a Molecular Theory of Electromechanical Responses}
\author{Keith A. Werling}
\author{Geoffrey R. Hutchison}
\author{Daniel S. Lambrecht}
\email{lambrecht@pitt.edu}
\affiliation{Department of Chemistry, University of Pittsburgh, 219 Parkman Avenue, Pittsburgh, PA 15260}

\begin{abstract}
  Developing a bottom-up (molecular) theory for the electromechanical response of aperiodic materials is a prerequisite
  for understanding the piezoelectric properties of systems such as nanoparticles, (non-crystalline) polymers, or biomolecule
  agglomerates. 
  The focus of this publication is to establish a new language and formalism for describing {\em molecular} 
  piezoelectric responses. 
  More specifically, we define the molecular piezoelectric response tensor \dd{}, which
  necessarily differs
  from the known bulk definition due to the anisotropy and inhomogeneity at the molecular scale, and derive
  an analytical theory to calculate this tensor. Based on this new theory, we develop a computational procedure for
  practical calculations of piezoelectric matrices for molecular systems. 
  Our studies demonstrate that the new analytical theory yields results that are consistent with fully numerical
  computations.  
  This
  publication is the first in a series; 
  this work establishes the theoretical molecular foundation and  
  follow-up publications will show how to bridge molecular and macroscopic
  piezoelectric responses. 
  It is expected that the present work will aid in developing design strategies for piezoelectric materials by revealing
  connections between molecular structure and piezoelectric response. 
  We expect that the language and formalism developed here may also be useful to describe mechanochemical phenomena. 
\end{abstract}

\maketitle

\section{Introduction}

The conversion between mechanical and other forms of energy is a ubiquitous process in many science and engineering applications.
Most notable from an energy footprint perspective is that a significant fraction of energy is inevitably lost by dissipating heat or vibrations (mechanical or acoustic).
Piezoelectric harvesting of mechanical into electric energy is therefore an interesting approach to reducing the energy
footprint of many applications that could for example help extend the battery life times of mobile devices or even
create self-powered electronic devices. 
Going in the opposite direction, generating mechanical deformations in response to electric stimuli
has multiple interesting applications such as for nano-actuators (e.g. piezo-motors in atomic force microscopy) or haptic feedback for microsurgery and other health related fields \cite{Dagdeviren2014}.
There is an ongoing interest to create organic piezoelectric materials to replace inorganic ceramics\cite{Werling0,Werling1, quan,Moody}.  
Organic materials offer several potential advantages, for example widely tunable mechanical properties [Matyjaszewski] and avoiding rare or
problematic elements such as lead or niobium.
Recently, polymer foam-based organic piezo-materials were reported with large piezoelectric responses (244 pC/N) \cite{Moody}, 
which demonstrates that it is possible to design organics that could replace inorganic ceramics. 

To realize the full potential of organic piezo-materials, it is necessary to develop
rational design approaches for organic piezoelectrics. 
Rationalizing organic polymer based piezoelectrics is very different from crystalline (bulk) materials in that it requires 
a multi-scale approach ranging from the (single) molecule limit to the bulk material. 
However, there is no established formalism, let alone an established language, for {\em molecular} piezoelectricity. 
For example, the existing theory of piezoelectricity is based on notions from continuum mechanics (bulk scale), to define quantities such as strain and stress,
typically derived for crystalline (periodic) systems based on unit cells and their deformations. 
This approach is not able to describe the piezoelectric response of individual molecules and, consequently, their
connection to the bulk response. 
Developing a molecular understanding of piezoelectric responses is, however, important
to describe (noncrystalline) polymers, as well as responses of small (e.g. nano) systems that are inhomogeneous,
anisotropic, and aperiodic, so that the bulk description is not yet applicable. 
Examples of the latter include the electromechanical response of biomolecules as well as nano-scale
machines.  
In this publication, we aim to establish a formalism and a language for {\em molecular} piezoelectric responses.
Establishing such a theory of molecular piezoelectric response is essential to being able to develop a bottom-up
(molecular-scale) understanding of piezoelectric responses in molecular materials. 
Furthermore, we also underline differences between {\em molecular} piezoelectric response and bulk response, 
emphasizing the need for a theory that can be applied and compare potential organic piezoelectric systems.


In our previous publications, we developed a computational approach to predict piezoelectric responses 
based on the definition of the converse piezoelectric effect, the deformation of a system in
response to an applied field. 
This approach is most natural in aperiodic calculations, where it is straightforward to apply an
electric field perturbation to the Hamiltonian. 
We established a simple computational procedure where the piezo-coefficient \dtt{} is
estimated by calculating the geometric response while applying a finite electric field. 
We then improved on this approach by developing an analytical expression that calculates \dtt{} from the zero-field 
geometric Hessian and dipole moment derivative. 
This approach has several advantages over our first procedure, for example by avoiding the
finite-field calculations (which can be problematic because of electronic instability of molecules in finite electric
fields) \cite{Yamabe1977, raey, raey1,raey2} and by cutting down on the computational cost via requiring only zero-field calculations (as opposed to
calculations at several finite field strengths). 
We demonstrated that both approaches yield comparable results for
molecular systems and used these approaches to explore piezoelectric responses in hydrogen-bonded systems. 
In fact, we showed that hydrogen bonding in 2-methyl-4-nitroaniline (MNA) gives rise to significant
piezoelectricity \cite{Werling1}. 
We then explored several examples of hydrogen donor-acceptor systems to help establish a general
rationale for the construction of systems with large \dtt{}. 
We found that our analytical expression is also useful in explaining piezoelectric responses by showing that a large \dtt{} requires both a large 
inverse Hessian (large compliance of the bond) and large dipole moment derivative. 
Point out papers by other authors that build on our approach / explore similar systems. 

The previously developed mathematical model exploited several circumstances to optimize the approach for the description
of piezoelectric responses in hydrogen-bonded systems. 
Firstly, we tacitly assumed that the hydrogen bond length maps onto the
deformation that one would observe at the bulk scale. 
In other words, we only took deformation along the H-bond axis into account.  
Secondly, we assumed that relaxation of the monomers within the field were negligible and would not
drastically affect the hydrogen-bond length \cite{Werling1}.  
We also exclude any anisotropic effects of the applied field on the hydrogen-bond length by
applying the field in only one direction.  
These assumptions allowed us to drastically reduce the dimensionality of the problem and map it to a single
variable $z$, the hydrogen-bond distance, which gave rise to the majority of
the piezoelectric response. 
These assumptions were backed by calculations that
show that the intramolecular piezoelectric response is at least an order of
magnitude smaller than the intermolecular response \cite{Werling1}.  
The mapping of the hydrogen bond-length to the \dtt{} works thanks to crystal packing effects that
confine the majority of the response to one plane. 
In these respects the potential of this model to describe the piezoelectric
effect for general types of organic crystals is rather limited indeed. 
Firstly, it is in general not guaranteed that there is a single, dominating direction of
largest piezoelectric response and even if there is, it might not be
straightforward to determine this direction a priori. 
In fact, in a sense molecules give rise to the most anisotropic cases imaginable compared to the
rather symmetric periodic cells.  
Therefore, we aim to extend our previous
approach by taking the full response of the nuclear coordinates into account.
This would reveal the full anisotropy in the response of the system and also does not make any assumptions about coordinates of the largest response.
We will show that the axis and magnitude of of the largest piezoelectric response for a given pair of coordinates in a system can be obtained 
by diagonalizing a matrix derived from the yet to be presented piezoelectric matrix.  
Furthermore, we derive a useful formula to calculate the piezoelectric matrix and discuss its connection to continuum strain mechanics. 
We will also underline the difficulty and impracticality of calculating a full 3rd rank piezoelectric tensor for molecules or small systems
while outlining connections of continuum strain theory to molecular deformations.  
The method we present for calculating the piezoelectric matrix can be used to understand the deformation properties between any two bodies in a system
and has great applicability to rational design of organic piezoelectric materials.
Another advantage is that this generalized approach can be automated and only requires that the user specifies the orientation of the 
system on which they wish to perform the calculation so that they may interpret the results relative to the systems orientation.
We discuss this and many other facets which users might find useful to understand and predict the response of molecular deformation in the presence
of fields.


The present publication is the first in a series aiming at the development of a bottom-up understanding of
electromechanical responses in aperiodic (inhomogeneous and anisotropic) systems. 
In this publication we will develop a description of the miscroscopic (single-molecule) electromechanical response. How to  
(Not supposed to be rigorous calculation of \dd{} to be compared to a bulk
measurement, but rather means of comparison for responses between individual
molecules. This is an area that does not have a language or formalism yet, but
requires one to quantify and rationalize. This derivation interfaces with
multiple relevant areas. 
Of course it would be extremely informative if one
could develop an understanding of how molecular piezoelectric response (which
we are treating here) maps onto the response as measured at the bulk scale.
Making this connection will be subject of future work. However, we expect that
developing this quantitative language of molecular piezoelectric response will
be enormously important for analyzing nano-scale (single-molecule) machines. It also 
connects well with the emerging area of mechanochemistry and we believe that
our approach will help lend mechanochemistry a quantitative language. 

This paper is organized as follows. In section II, we provide a brief background of strain theory and the linear
piezoelectric equation in the established language of bulk deformations. Since we feel that many chemists are unfamiliar
with continuum mechanics, we provide a more complete introduction to strain theory in appendix A and derivation of the Green strain tensor 
in the connection to longitudinal strain. Section III presents
our derivation of molecular piezoelectric response, which is split into five parts. 
In section III.A we simplify the equations for the field derivative of longitudinal strain that we introduce in the background section.
In section III.B we indrocue the piezoelectric matrix and make connections from bulk strain theory to 
molecular systems.
In section III.C we describe how to construct the displacement field derivative for molecules and the 
nuance of how to project out unwanted motions.
In section III.D we show how to calculate the piezoelectric matrix for molecules and some ways it can be used
to screen for ``good'' organic piezoelectric candidates.
In section III.E we describe a rank-4 piezoelectric tensor which for a molecule takes the place of the 
rank-3 piezoelectric tensor field in a bulk system.
At points necessary in section III, we discuss how deformation response in molecules is uniquely distinguished 
from bulk response and the inherent anisotropy and inability to ascribe a full piezoelectric system to molecular systems.
Section IV outlines the computational procedure for calculating molecular piezoelectric responses. 
In section V we present numerical results for molecules of interest to illustrate the concepts derived in the previous
sections. 
We conclude with a discussion and outlook in section VI. 

\section{Background: Introduction to Piezoelectricity}


Piezoelectricity is a phenomenon that relates mechanical deformation with charge separation and vice versa.  Piezoelectric compounds are ubiquitous in nature (bone, collagen, proteins, crystals, etc.) and 
industry (zinc oxide, lead zirconate titanate, etc.).  
When piezoelectric materials are compressed or deformed in some
manner, they become polarized and the difference in chemical potentials at opposite surfaces can be measured
as a voltage difference.  
This process is due to either the reorientation of dipoles and their corresponding Weiss domains (local areas of similar dipole density) under applied stress or a local 
change in the environment surrounding the domains.
Conversely, piezoelectric materials are also deformed in response to an applied electric field. This deformation is due
to the field interacting with the molecular electrostatic moments (both static and induced). 
For the purposes of this manuscript, we will be concerned with the linear piezoelectric equations, which pertain to
infinitesimally small applied stresses and electric fields, respectively.   
The theory of piezoelectric response is well-established for crystalline (bulk) systems. 
For more information on piezoelectricity we refer the reader to Ref. \cite{ieeePiezo,Martin1972, Cohen:2008gh}
The {\em direct} piezoelectric effect describes the polarization (electric charge density displacement) $D_{i}$ of the crystal in terms of an applied
stress $T_{jk}$ and electric field $f_j$, 
\begin{equation}
\begin{aligned}
\label{eq:direct}
D_i = d_{ijk}T_{jk}+\epsilon_{ij}f_j \\
\bf{D} = \bf{dT}+\bf{\epsilon f}.
\end{aligned}
\end{equation}
Here and in the following, Einstein sum convention is implied for repeated indices. 
The {\em converse} piezoelectric effect describes the strain $E_{ij}$ 
induced in the material when an external field is applied, 
\begin{equation}
\begin{aligned}
\label{eq:convers}
E_{ij} &= \sigma_{ijkl}T_{kl}+d_{ijk}f_k \\
\bf{E} &= \bf{\sigma T}+\bf{df}. 
\end{aligned}
\end{equation}
Here the tensor $\bf{E}$ is known as Green strain tensor. 
The stress tensor, $\bf{T}$, is linked to the strain tensor
by the rank-4 compliance tensor, $\bm{\sigma}$, at zero field. 
For aperiodic systems it is easiest to work with the converse equations by applying an electric field without any
stress. 
%
The piezoelectric tensor is then given by 
\begin{equation}
\label{eq:piezoTen}
d_{ijk} = \left(\frac{\partial E_{ij}}{\partial f_k}\right)_{{\bf T} = {\bf 0}}.
\end{equation}
We note that the indices pertaining to strain, ${ij}$, depend on the coordinate system used as a reference frame
for measuring strain (which can be e.g. unit cell parameters or cartesian coordinates, etc.), whereas the index contracted with the
field, $k$, depends on the external (laboratory) reference frame of the applied field.

To understand how to calculate $d_{ijk}$ in practice and to provide some context for the following sections, we 
introduce here selected concepts of strain theory within the framework of continuum mechanics. 
In continuum mechanics, the material is treated as a   
collection of a continuum of infinitesimally small particles. 
Obviously, this is not an appropriate description for a
molecular system. 
Starting from these definitions, we will therefore develop a theory for molecular (i.e., discrete, anisotropic and inhomogeneous) systems. 
A more complete derivation of the Green strain tensor can be found in the Appendix. 
Our discussion closely follows the presentation from Ref. \cite{contMech}.

\begin{figure}
  \centering
  \includegraphics[width=0.48\textwidth]{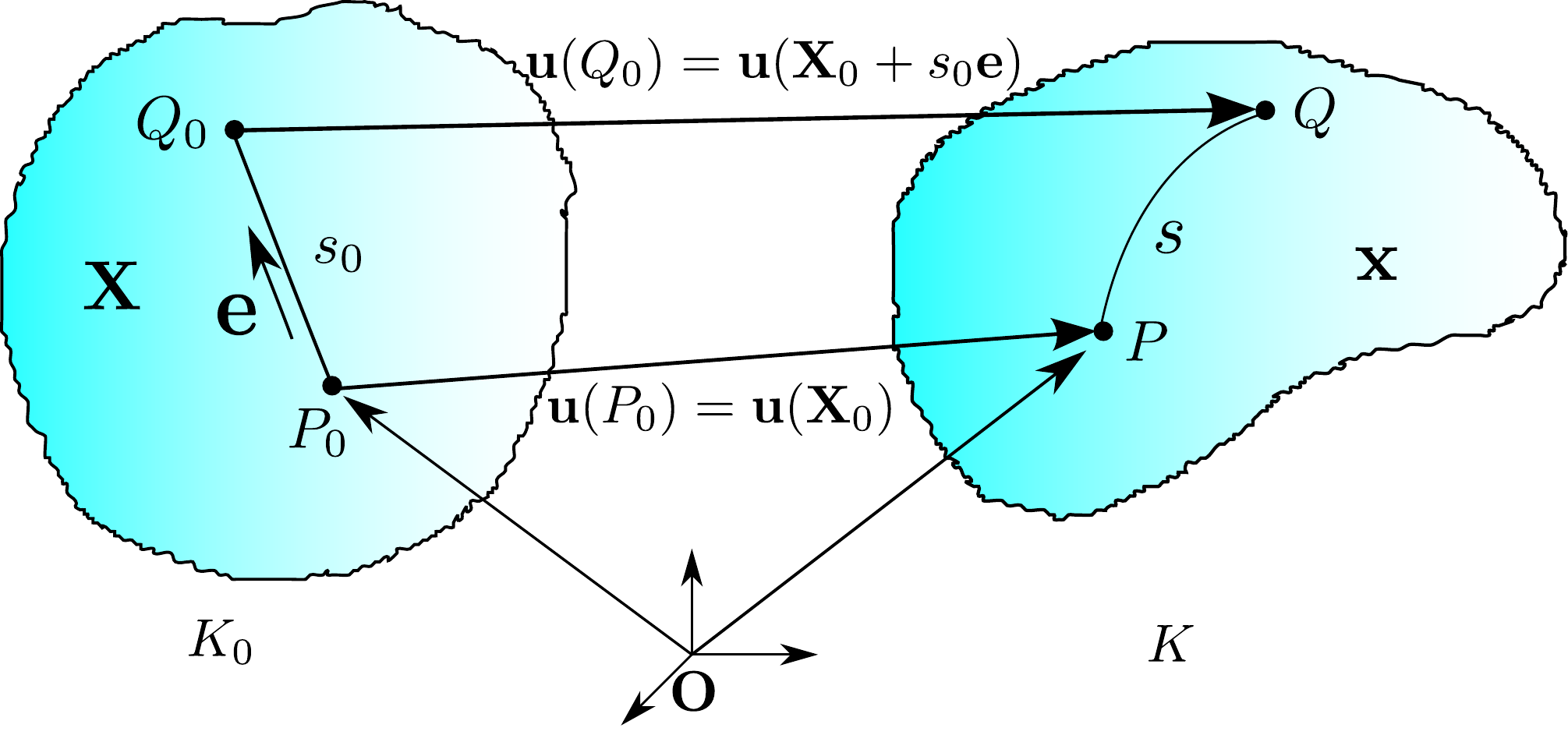}
  \caption{General deformation of a body. Adapted from Ref. \cite{contMech}. The position vectors for the undeformed body, $K_0$, and the deformed body, $K$ are given by ${\bf{X}}$ 
  and ${\bf{x}}$ respectively.  The displacement vector, ${\bf{u}}$ map the position in the undeformed coordinates to a position in the deformed coordinates, and indicate the direction and 
  magnitude in the ``shift'' of an infinitetessimal particle in space as the body deforms.  }
  \label{fig:deform}
\end{figure}

Generally, strain analysis is concerned with concepts such as longitudinal strain, shear strain, and volumetric strain.
We will be mostly concerned with longitudinal strain for the discussion that follows.
To this end, consider a body undergoing mechanical deformation (\fig{fig:deform}).
Configurations $K_0$ and $K$ correspond to the undeformed and deformed bodies, respectively. 
Particle positions before and after deformation are denoted by $X$ and $x$, respectively, where we note that these
positions are specified within the same (external) coordinate frame. The vector field ${\bf{u}}({\bf{X}}, t)$ gives the displacement vector from a point the undeformed body 
to a corresponding point in the deformed body.  The displacement vector field will become important later when we define the Green strain tensor.  For molecular systems the 
a discrete displacement vector field exists for each atom in a molecule or a system as said body deforms.  
A central question is how a given material line deforms as the body is deformed from the initial ($K_0$) to the final
($K$) shape. 
For our purposes, the material line can be seen as any line drawn through some arbitrary but continuous path of particles
in the body. For convenience, we show the undeformed material line as a straight line segment $P_{0}Q_{0}$ with length
$s_0$ and the deformed material line as the curve along $PQ$ with length $s$. 
The unit vector along the line segment $P_{0}Q_{0}$ is called $\bld{e}$. 
Based on this picture, one defines the longitudinal strain $\epsilon$, i.e. the strain along the direction of $\bld{e}$,
as the relative change in the length of the material line upon deformation in the limit of infinitesimal line length:
\begin{equation}
\label{eq:longstrain}
\epsilon=\lim_{s_0\to 0}\frac{s-s_0}{s_0}=\frac{ds-ds_0}{ds_0}=\frac{ds}{ds_0}-1
\end{equation}
We note that we used a very similar, though less rigorous definition in our previous papers \cite{Werling0,Werling1} where we used the
percent deformation in bond lengths to approximate the piezoelectric deformation.   
As noted, we are not concerned with volumetric or shear strain here, although the Appendix presents more details about
the Green strain tensor which can be used to completely specify the deformation around a point within a body.

The goal now is to find working equations for calculating the longitudinal strain $\epsilon$ from \eq{eq:longstrain}. 
To this end, one needs to determine the arc length $s$ of the deformed material line. 
Treating $s_0$ as a variable parameter, any point along the undeformed material line $P_{0}K_{0}$ can be
written as $\bld{X}_{0} + s_0 \cdot \bld{e}$, where $\bld{X}_0$ is the starting point of the undeformed material line. The positions
of particles in the deformed bodies can then be defined as functions of the undeformed point as well as a time
coordinate $t$ that determines the progress of the deformation: $\bld{x} = \bld{x}(\bld{X}_{0} + s_0 \cdot\bld{e}, t)$.
The arc length of $s$ as a function of $s_0$ (which we treat as a curve parameter) is then  
\begin{equation}
\label{eq:arclen}
s(s_0) = \int_0^{s_0}\sqrt{\frac{dx_i}{d\bar{s}_0}\frac{dx_i}{d\bar{s}_0}}d\bar{s}_0, 
\end{equation}
which suggests that the arc length derivative with respect to $s_0$ and it's square are
\begin{equation}
\label{eq:dsds0}
  \begin{split}
    \frac{ds(s_0)}{ds_0} &= \sqrt{\frac{dx_i}{ds_0}\frac{dx_i}{ds_0}} \\
     \bigg(\frac{ds}{ds_0}\bigg)^2 &= \frac{dx_i}{d\bar{s}_0}\frac{dx_i}{d\bar{s}_0}
  \end{split}
\end{equation}
The derivative of the deformed arc length with respect to the undeformed length parameter is more easily cast in
terms of a mixed derivative of the deformed coordinates $\bld{x}$ with respect to the undeformed coordinates $\bld{X}$,
\begin{equation} 
\label{eq:dGrad}
  F_{ik} = \frac{\partial x_{i}}{\partial X_{k}},
\end{equation}
to yield 
\begin{equation}
\label{eq:dsds022}
 \left(\frac{ds}{ds_0}\right)^2
 = \bld{e} \cdot (\bld{F}^{T} \bld{F}) \cdot \bld{e} \\
 \end{equation}
 $(\bld{F}^{T} \bld{F})$ is known as the Green deformation tensor and which we denote by ${\bf{C}}$.  
 Appendix A shows the derivation for the relationship between the Green deformation tensor and the Green Strain tensor, ${\bf{E}}$, but we arrive at  
 \begin{equation}
 \begin{split}
 \left(\frac{ds}{ds_0}\right)^2 = \bld{e} \cdot \bld{C} \cdot \bld{e} \\
 &\equiv \bld{e} \cdot (\bld{1} + 2\bld{E}) \cdot \bld{e}  \\
 &= 1 + 2\bld{e} \cdot \bld{E} \cdot \bld{e},
\end{split}
\end{equation}
using $C = \bld{1} + 2\bld{E}$.
With these definitions, we arrive at a working equation for longitudinal strain along a material line defined by the
unit vector $\bld{e}$ as 
\begin{equation}
\label{eq:longstrain2}
  \epsilon 
  = \frac{ds}{ds_0} - 1
  = \sqrt{ 1 + 2{\bf{e}}\cdot \bf{E} \cdot \bf{e}} - 1
\end{equation}
In the limit of small deformations, the working equation for the longitudinal strain becomes
\begin{equation} 
\label{eq:small}
  \epsilon = \bld{e} \cdot \bld{E} \cdot \bld{e}
\end{equation}
which can be seen from a taylor expansion about the point $\bld{e} \cdot \bld{E} \cdot \bld{e} = 0$
We have not yet defined the elements of the Green strain tensor, ${\bf{E}}$, but they are given by 
\begin{equation}
\label{eq:gst}
 E_{kl}=\frac{1}{2}\bigg(\frac{\partial u_k}{\partial X_l}  + \frac{\partial u_l}{\partial X_k} +
  \frac{\partial u_i}{\partial X_k}\frac{\partial u_i}{\partial X_l}\bigg)
\end{equation}
and is derived fully in the Appendix.  
The Green strain tensor only depends on the derivatives of the displacement vector field with respect to the 
undeformed position vectors. Just as the displacement is a vector field with a unique vector specified for every point in the undeformed body, the Green strain tensor 
is a tensor field that fully specifies the strain around a given point in the deformed body in reference to the undeformed body.  

For convenience later on, we may define the matrix ${\bf{A}}$, and recast the Green strain tensor in terms of ${\bf{A}}$.
\begin{equation}
  \label{eq:A}
  A_{ik} = \frac{\partial u_i}{\partial X_k},
\end{equation}
 and recast the Green strain tensor in terms of ${\bf{A}}$.
 \begin{equation}
 \label{eq:H2E}
  {\bf{E}}= \frac{1}{2} \left( \bld{A} + \bld{A}^{T} + \bld{A}^{T}\bld{A} \right),
\end{equation}
We now wish to show how we can approximate \dtt in a manner equivalent to our previous papers \cite{Werling0,Werling1}.
For our situation, we take the undeformed body, $K_0$ to be the optimized geometry at zero field.  
The only source of deformation for our system will be the applied electric field ${\bld{f}}$, which implies that the Green strain tensor
is also a function of ${\bld{f}}$, i.e. $\bld{E} = \bld{E}(\bld{f})$.  We can therefore rewrite \eq{eq:longstrain2}.
\begin{equation} 
  \epsilon = \sqrt{1 + 2 \, \bld{e} \cdot \bld{E}(\bld{f}) \cdot \bld{e}} - 1
                \approx \bld{e} \cdot \bld{E}(\bld{f}) \cdot \bld{e}
\end{equation}
If we take the derivative with respect to $\bld{f}$ we obtain the following.
\begin{equation}
\label{eq:longstrainDeriv}
\frac{\partial \epsilon ({\bf{f}})}{\partial {\bf{f}}} = \frac{{\bf{e}}\cdot \frac{\partial {\bf{E}}({\bf{f}})}{\partial {\bf{f}}} \cdot {\bf{e}}}{\sqrt{ 1 + 2{\bf{e}}\cdot \bf{E(f)} \cdot \bf{e}}}
\end{equation}
Because we choose $K_0$ as our system at zero field, $\bld{E}(\bld{f}=\bld{0})=\bld{0}$ ($\bld{0}$ on the right hand side is a matrix not a vector)--ie. the displacement derivatives contained
in $\bld{A}$ are all zero since the $\bld{u}=\bld{0}$ everywhere. 
If we evaluate \eq{eq:longstrainDeriv} at $\bld{f}=\bld{0}$ the derivative reduces to the following equation.    
\begin{equation}
\label{eq:longstrainDeriv2}
\frac{\partial \epsilon ({\bf{0}})}{\partial {\bf{f}}} = {\bf{e}}\cdot \frac{\partial {\bf{E}}({\bf{0}})}{\partial {\bf{f}}} \cdot {\bf{e}}
\end{equation}
\eq{eq:longstrainDeriv2} not surprisingly matches the field derivative of the longitudinal strain at small defomations (\eq{eq:small}). 
The derivative of the Green strain tensor $\bld{E}$ with respect to the field, $\bld{f}$ is just the piezoelectric tensor, $\bld{d}$ (\eq{eq:piezoTen}). 
Thus the derivative of the longitudinal strain at zero field reduces to the following:
\begin{equation}
\label{eq:longstrainDeriv3}
\frac{\partial \epsilon ({\bf{0}})}{\partial {\bf{f}}} = {\bf{e}}\cdot \bld{d}({\bf{0}}) \cdot {\bf{e}}
\end{equation}
In the sections to come, we will reduce \eq{eq:longstrainDeriv3} to a simpler form, devise a method for calculating piezoelectric coefficients for molecules,
and introduce the piezoelectric matrix, $\bld{P}$.  \eq{eq:longstrainDeriv3} can be used to approximate \dtt in the same manner as our previous papers \cite{Werling0,Werling1}, but is a generalization, which
we will find useful later.  Furthermore, it connects the work we have done thus far to strain theory, which we hope serves as a tool for future work in this area.

\section{Derivation of a Theory of Molecular Piezoelectric Response}

In our previous publications \cite{Werling0,Werling1}, we described two simple approaches to calculate molecular piezoelectric responses in
systems such as hydrogen-bonded organic crystals. 
This type of system is relatively simple to describe, since the
dominant contribution to the piezoelectric response arises from the hydrogen bond. 
Consequently, our previous approaches
focused on predicting, either numerically \cite{Werling1} or analytically \cite{Werling0}, the piezoelectric response along a single
coordinate (in this case, the hydrogen bonding coordinate).  
Our first publication used the most simple conceivable
approach, applying finite electric fields along the hydrogen bond axis, and measuring the geometric deformation as a
function of field strength. 
This approach is problematic in that, strictly speaking, static electronic structure
calculations in finite fields are not well-defined due to the instability of the molecule in the field. 
It furthermore
requires a relatively large number of geometry optimizations at various field strengths.  
The second approach \cite{Werling0} used
an analytical expression for \dtt{} derived from a Taylor expansion around zero field strength and zero displacement.
This approach is therefore well-defined from an electronic structure point of view, and it leads to a significant
reduction of computational cost because it requires only one geometry optimization as well as the curvature along a 1D
potential curve.  
Naturally, this approach is restricted to predict only a single contribution to the full piezoelectric
tensor at a time, where one would typically choose the coordinates so that the largest contribution to the
piezo-response is obtained (\dtt{}). 
Naturally, the approach is rather unwieldy to use for systems where one cannot
identify, a priori, an individual coordinate to describe the piezoelectric response. 
However useful for hydrogen-bonded piezoelectrics, the simple 1D approach also neglects the coupling between deformations 
along different coordinate axes. 
For these reasons, the present section aims to develop a complete formalism for the description of molecular
piezoelectric responses, irrespective of the dimensionality of the potential energy surface and independent from any
assumptions about preferred axes for the response. 
This new approach will help extend the theory of piezoelectric response to arbitrary classes of molecules. 
Examples of
systems that we are particularly interested in are single-molecule responses in organic systems such as helicines or
peptides. 
To this end, we develop a language for molecular electromechanical responses based on, and making connections to, strain
theory as known from continuum mechanics. 
While aimed at deriving definitions and working equations for practical
applications, this work touches on some fundamental aspects of the philosophy of science, namely the question at what
size it is appropriate to describe a system as (continuum) material and when as a molecule, and how to reconcile the
languages of both scales \cite{hartmann}. 
This approach will take into account the fully-dimensional relaxation of the
system in response to an applied electric field. 
As such, our approach is valid for any direction of piezoelectric
responses (or combinations thereof), and does not require any a priori knowledge of the preferred (dominant) axes. 
This
feature is essential to enable automating the process of calculating molecular piezoelectric responses for applications such
as computational screening. 

\subsection{Simplifying the Equation for the Piezoelectric Coefficient}

Recall \eq{eq:longstrainDeriv3}.  
We have claimed that this equation is equivalent to our previous methods for estimating \dtt for molecules \cite{Werling0}. 
Before we can make this connection, we will find it useful to first simplify \eq{eq:longstrainDeriv3}.  
We can reduce \eq{longstrainDeriv2} further by using the definition for the Green strain tensor $\bld{E}$ (\eq{eq:H2E}). 
We first rewrite \eq{eq:longstrainDeriv2} elementwise (the derivative is a vector quantity with components of field) using the definition of $\bld{A}$ (\eq{eq:A}).
\begin{equation}
\begin{split}
\label{eq:element}
\frac{\partial \epsilon ({\bf{0}})}{\partial f_l}&= \frac{1}{2}e_i \bigg(\frac{\partial^2 u_i(\bld{0})}{\partial X_k\partial f_l} + \frac{\partial^2 u_k(\bld{0})}{\partial X_i \partial f_l} + \\
 &\frac{\partial^2 u_j(\bld{0})}{\partial X_i \partial f_l}\frac{\partial u_j(\bld{0})}{\partial X_k} + \frac{\partial u_j(\bld{0})}{\partial X_i}\frac{\partial^2 u_j(\bld{0})}{\partial X_k \partial f_l}\bigg)e_k
 \end{split}
\end{equation}
The last two terms in \eq{eq:element} result from the product rule for $\bld{A}^T\bld{A}$ from \eq{eq:H2E}.  
Both terms vanish when evaluated at zero field.  
This is because $\frac{\partial u_j(\bld{0})}{\partial X_i}=A_{ji}(\bld{0})=0$ (every element is zero) if
we take the zero field body as $K_0$ and $K$ (which is equivalent to evaluating $\bld{A}$ at zero field. 
If this is the case, the displacement vector for every point in the body is constant (and actually $\bld{0}$), and hence 
the derivative in any direction ($X_i$) vanishes (see Appendix for more information on strain theory).  
We can therefore rewrite \eq{eq:element}, ignoring the last two terms.  
\begin{equation}
\label{eq:element2}
\frac{\partial \epsilon ({\bf{0}})}{\partial f_l}= \frac{1}{2}e_i \bigg(\frac{\partial^2 u_i(\bld{0})}{\partial X_k\partial f_l} + \frac{\partial^2 u_k(\bld{0})}{\partial X_i \partial f_l} \bigg)e_k
\end{equation}
We can rewrite \eq{eq:element2} in terms of matrix derivatives for simplicity.  
\begin{equation}
\begin{split}
\label{eq:nonelement}
\frac{\partial \epsilon ({\bf{0}})}{\partial \bld{f}}&= \frac{1}{2}\bld{e} \bigg(\frac{\partial \bld{A}^T(\bld{0})}{\partial \bld{f}} + \frac{\partial \bld{A}(\bld{0})}{\partial \bld{f}} \bigg)\bld{e} \\
\frac{\partial \epsilon ({\bf{0}})}{\partial \bld{f}}&= \bld{e}\frac{\partial \bld{A}(\bld{0})}{\partial \bld{f}}\bld{e}
\end{split}
\end{equation}
Because both matrices are contracted by $\bld{e}$ on both sides, and one is the transpose of the other in these two dimensions, both terms in the first line were equivalent and reduced to twice
the value of one term, which cancelled the prefactor of $\frac{1}{2}$ in the final line.
$\frac{\partial \bld{A}(\bld{0})}{\partial \bld{f}}$ is a 3 dimensional tensor, and it is equivalent to the field derivative of the Green strain tensor evaluated at zero field (ie. the piezoelectric tensor evaluated
at zero field).  
We will work with \eq{eq:nonelement} in the sections to come.  

\subsection{Molecular connections to Strain Theory and the Piezoelectric Matrix}

Thus far, we have claimed that the derivative of the longitudinal strain equation (eqn) with respect to the applied electric field system
is related to our calculations of \dtt in our previous work \cite{Werling0,Werling1}.  
In this section we will attempt to explain geometrically the connection of strain theory to our work, and apply our equations to molecular 
systems.
Furthermore, we will demonstrate why, for molecules, the full piezoelectric tensor is untenable for calculation and introduce the piezoelectric 
matrix $\bld(M)$, which will take the place
of the piezoelectric tensor for molecules.  

In the Background section we presented a working equation (\eq{eq:longstrainDeriv3}) with dependence on the piezoelectric tensor ($\bld(d)$),
and we further reduced this equation to \eq{eq:nonelement} for the 
zero field evaluation of \dtt.  
We have yet, however, to explain the direct connection of \eq{eq:nonelement} to our previous work.  
To interpret \eq{eq:nonelement} geometrically, we need to look back to \fig{fig:deform}.  
As we recall, longitudinal strain deals with how material lines in continuum bodies deform.  
The direction of the material line in the undeformed
body was given by $\bld{e}$. 
In \eq{eq:nonelement}, two of the indexes of $\frac{\bld{A}(\bld{0})}{\partial \bld{f}}$ are contracted with $\bld{e}$--namely the indexes of the 
undeformed coordinates $X_i$ and displacement vector coordinates $u_i$.  
If we ask what the contractions with $\bld{e}$ mean, we can gain insight into how to understand \dtt and furthermore connect strain theory to 
molecular piezoelectricity and point out differences.  
First we define a vector $\bld{X}$ in the direction of $\bld{e}$ with a length parameter of $s_0$ (equivalent to that shown in 
the undeformed body in \fig{fig:deform}).  
\begin{equation}
 \bld{X} = s_0 \bld{e}
\end{equation}
Taking the derivative with respect to $s_0$ tells us how the vector changes per unit change in the length of $s_0$
\begin{equation}
\label{eq:dxds_0}
 \frac{d\bld{X}}{ds_0}= \bld{e}  \equiv \frac{dX_i}{ds_0} = e_i
\end{equation}
Here we have a useful definition of $\bld{e}$.  
The change in coordinates in the undeformed system per unit of $s_0$ corresponds to the 
vector $\bld{e}$, which gives the direction of our material line in the undeformed body.  
With this definition we may further simplify \eq{eq:nonelement}.  We will
again use the definition of $\bld{A}$.  
\begin{equation}
\begin{split}
\label{eq:Meq}
\frac{\partial \epsilon ({\bf{0}})}{\partial f_l}&= e_i\frac{\partial^2 u_i(\bld{0})}{\partial f_l \partial X_k}e_k = e_i \frac{\partial^2 u_i(\bld{0})}{\partial f_l \partial X_k}\frac{dX_k}{d_{s_0}}\\
\frac{\partial \epsilon ({\bf{0}})}{\partial f_l}&=e_i\frac{\partial^2 u_i(\bld{0})}{\partial f_l \partial s_0}\\
\frac{\partial \epsilon ({\bf{0}})}{\partial f_l}&=e_i P_{il}(\bld{0})\\
\frac{\partial \epsilon ({\bf{0}})}{\partial \bld{f}}&=\bld{e}\bld{P}(\bld{0})
\end{split}
\end{equation}
We have substituted in for $e_k$ our definition in \eq{eq:dxds_0} and contracted over the index $k$ in the zero field piezoelectric tensor, 
corresponding to the undeformed coordinates $X_k$. 
We have renamed the resulting matrix from this contraction $\bld(P)=\frac{\partial^2\bld{u}}{\partial s_0 \partial \bld{f}}$.  
We will call ${\bf{P}}$, in the sections to come, the piezoelectric matrix.  
Here we focus on the matrix in the context of a continuum body; later we will apply the ideas to discrete molecular systems.
As we move along the material line in the undeformed coordinates, $s_0$ is the curve parameter for the displacement vectors to the deformed coordinates (see \fig{fig:deform}).  
We may evaluate the vector derivatives of the the displacement vector $\bld{u}$ as we move infinitesimally along $s_0$.  
This gives us $\frac{\partial \bld{u}}{\partial s_0}$ evaluated
at our point of interest in the body.  
We may then evaluate the derivative of $\frac{\partial \bld{u}}{\partial s_0}$ with respect to the field vector at a 
given field magnitude (which in our case is $0$).  
We then obtain the matrix $\bld{P}$.  
Of course it is equally valid to view the matrix by looking at $\frac{\partial \bld{u}}{\bld{f}}$ at a given point in our body first and 
then asking how this matrix changes by moving infinitesimally along our material line.  
The order we take the derivatives of $\bld{u}$ should not matter because we assume 
$\bld{u}(\bld{f},s_0)$ to be smooth and obey Euler's rule for mixed derivatives.  

In the sections to come $\bld{P}$ will become very important for us.  
First, however, we must finish explaining the connection between \eq{eq:nonelement}, which we have just rewritten
as \eq{eq:Meq}.  
We need to explain the other contraction with $\bld{e}$, which contracts with the index for the displacement vector field, $\bld{u}$.  
Since $\bld{e}$ is a unit vector, if we project $\bld{u}$ onto $\bld{e}$ the resulting vector will have the length $\bld{u}\cdot \bld{e}$. 
We can name this length variable $v$, we recognize it is merely a linear combination
of the individual components of $\bld{u}$.  $
v$ describes the portion of $\bld{u}$ which is along $\bld{e}$, which is the direction of our undeformed material line.  
\begin{equation}
 v = \bld{u}\cdot \bld{e} = u_i e_i
\end{equation}
Hence, any derivatives of $\bld{u}$ will carry through in the typical manner for linear equations.
\begin{equation}
\label{eq:prevWork}
 \frac{\partial^2 v}{\partial s_0 \partial f_l} =  e_i \frac{\partial^2 u_i}{\partial s_0 \partial f_l}
\end{equation}
If we evaluate these derivatives at a particular point in the undeformed body and at zero field, we obtain a result identical to \eq{eq:Meq}.  
Hence, \eq{eq:Meq} describes field derivative of the change 
in the displacement vector along $\bld{e}$ per unit of change along the material line in the undeformed body.  
In simpler terms, as we move along the undeformed material line,
the piezoelectric matrix measures how the strain changes per unit of field in each direction.

From \eq{eq:prevWork}, which is equivalent to \eq{eq:longstrainDeriv3}, we can easily see how evaluating the field derivative of longitudinal strain is similar to our previous work \cite{Werling0,Werling1}.
In our previous work we estimated the deformation of organic dimer systems in the direction of an applied field.  
In this work we separated the dimers by aligning two atoms (usually hydrogen-bonded)
along a particular axis and estimated how applying a field in this direction would change the distance between these two atoms.  
We accomplished this through a potential energy ''scan``
along this separation coordinate.  
However, we only approximated the deformation in the direction of the original bond direction.  
For example, if the hydrogen-bond was along the z-axis,
we would only approximate the deformation of the bond-along the z-axis--even though there are three possible dimensions in which the bond may deform.  
\eq{eq:prevWork} is similar in that 
the contraction the components of ${\bld{u}}$ with ${\bld{e}}$, projects the deformation onto the original material line direction ${\bld{e}}$.  
The only difference is that the result of 
\eq{eq:longstrainDeriv3} is a vector with field components because the derivative was taken with respect to each field component.  
If we wanted to recover a similar approximation to \dtt, which
mirrors our previous work, we would either take the derivative of the longitudinal strain with respect to just one component of the field along 
the direction of the original bond, or of 
course we may project the result of \eq{eq:prevWork} onto a field length parameter in the direction of ${\bld{e}}$, 
in a manner akin to projecting the displacent derivative with respect to the undeformed
coordinates, ${\bld{X}}$, onto the parameter $s_0$.  

All of the work described so far has been done in the spirit of continuum mechanics.  
Meaning that on any length scale the system will always contain matter.  
For our molecular considerations in this work, the continuum hypothesis clearly does not hold.  
We are mostly interested in calculating something similar to piezoelectric coefficients for single or several molecular
systems that are in general aperiodic.  
In a periodic system like a crystal, we can talk about the deformation of unit cells in which case the 
cell dimensions act like inifinitesimal vectors which are the basis for the displacement derivatives discussed above.  
In this case, the continuum derivatives involved in strain are easily translated into the molecular realm.  
However, when we have single molecules or small systems which show marked anisotropic variety, we lack the unit cells or raw 
choice of basis vectors to describe the deformation of the 
system.  
It is for this reason that it is impossible to calculate the full third rank piezoelectric tensor for a molecule.  
If we were to ask how the molecule or system deforms as we move in the 
$x$, $y$,or $z$ direction, we would be at a loss for how to adequately describe the response.  

\begin{figure*}
  \centering

  \includegraphics[width=1.00\textwidth]{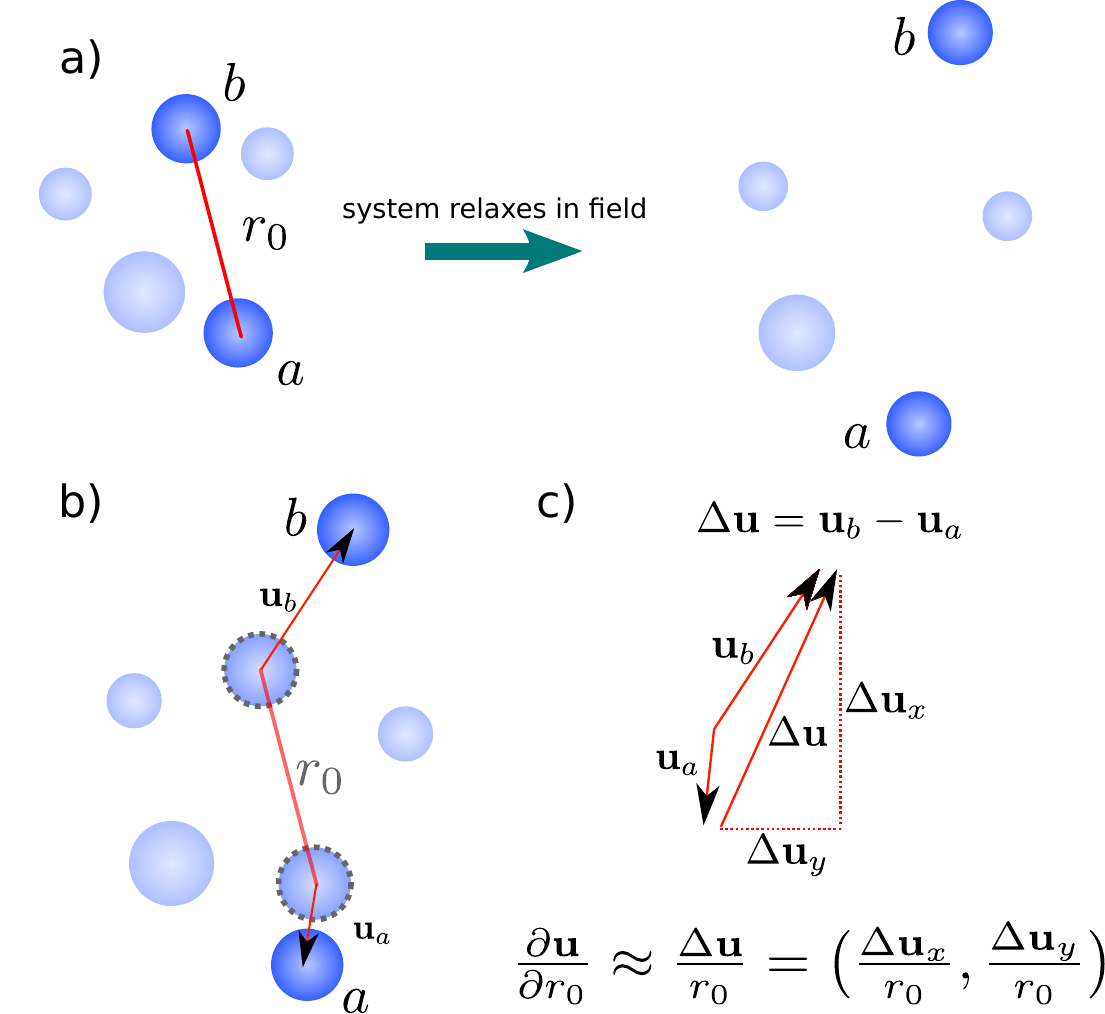}
  \caption{We draw a material line between nuclei a and b in a hypothetical small system.  A) If we were to apply a field, the nuclei would be allowed to ''relax`` within the field to find a minimum in energy. 
  B)We calculate the displacement vectors for each nuclei from their previous to current positions. C)}
  \label{fig:defAnalysis}
\end{figure*}
Consider the following example.
If we had the system dipicted in \fig{fig:defAnalysis}, we may draw what is, in the molecular sense, a material line.  
In the continuum sense a material line connects adjacent (in a continuous sense)
points of the undeformed body, and in general, if the body is deformed in a ''smooth`` way, we may draw a continuous line in the undeformed body representing the same 
set of points (as shown in \fig{fig:deform}).
For a molecule, there is no way to do this.  
There is space between the nuclei, or rather a sea of electron density.  
To make the connection to strain theory, we shall use the same 
idea presented in our previous work \cite{Werling0,Werling1}; we will choose a line drawn between any two nuclei in the system as a material line of interest.  
If this is the case, we may draw many ''material lines,`` and in some cases they will point in similar directions as other material lines, 
and could in fact overlap.  
If we look at the deformation in \fig{fig:defAnalysis}, we can indicate with vectors the relative deformation of 
two pairs of nuclei (which act as a material line).  
Even though these material lines occupy nearly the same region of space and are in a similar direction, they can have markedly different 
relative deformations (and hence piezoelectric matrices) between the pairs.  
In a continuum material, we assume that a single tensor can fully describe the piezoelectric properties of 
an infinitesimally small region of the material, and if we wish to determine the piezoelectric matrix for any direction, we would contract over the 
index for the undeformed body coordinates
with a directional vector in the direction of the new material line.  
For a molecule, if we take the approach presented here and use a pair of nuclei to define a material line, we cannot
construct a full piezoelectric tensor which adequately describes the whole molecule or even a part of the molecule.  
The best we could do is to write a piezoelectric matrix for every
pair of nuclei in the molecule or system because the tensorial properties of the molecule do not vary in a continuous 
manner but show extreme anisotropy to the point where deformation in 
similar directions may be very different (ie. there is no bijection connecting points in the system to a piezoelectric tensor).

One might wonder why we bother with strain theory at all if this were the case.  
We could afterall develop this theory from simple geometric arguments and nothing would change--we would merely dispense
with the connections to continuum body mechanics.  
The reason we approach this problem from the view point of continuum mechanics is so that we can use this formalism for any further development 
of this theory (time depenence, nonequilibrium and finite field calculations) and so that we may compare single molecule deformation to the full 
piezoelectric response of a crystal.

\subsection{The Molecular Displacement Derivative with Respect to Field}
We now wish to derive a practical way to calculate piezoelectric matrices for molecular systems.
As mentioned in the above section, when constructing the piezoelectric matrix $\bld(P)=\frac{\partial^2\bld{u}}{\partial s_0 \partial \bld{f}}$,
we may choose either to take the field derivative of $\frac{\partial \bld{u}}{\partial s_0}$ or the derivative of 
$\frac{\partial \bld{u}}{\bld{f}}$ with respect to the material line length parameter $s_0$.  
We will find the latter of the two methods advantageous because it requires no actual finite field calculations.
In a continuum body we would calculate $\frac{\partial \bld{u}}{\bld{f}}$ at some point of interest or as a tensor field for the whole body.
The molecular equivalent is to calculate $\frac{\partial \bld{u}}{\bld{f}}$ for all of the nuclei of the system--ie. here we take the vector
${\bf{u}}$ to be the displacement for every nuclear coordinate as a molecule or small system relaxes in a field.
Therefore, the matrix $\frac{\partial \bld{u}}{\bld{f}}$ will have dimensions of $3Nx3$.

In our previous paper we calculate the derivative of the hydrogen-bond length parameter with respect to the field magnitude in the z direction
$\frac{dz}{df}$.
The derivation will proceed via a similar routine for $\frac{\partial \bld{u}}{\bld{f}}$ but adds the complication of 
multiple variables, and as we shall see, rotation and translation contamination.  
To this end, we use a Taylor series of the molecular energy in terms of atomic displacements \uu{} for an arbitrary molecular system 
and electric field vector \ff{},
\begin{equation}
\begin{aligned}
E(\uu,\ff) 
  &= E(0,0) + \bld{g}_{\uu}^{T} \uu + \bld{g}_{\ff}^{T} \ff + \frac{1}{2} \, \uu^T \bld{H}_{\uu\uu} \uu \\
  &+ \frac{1}{2} \, \ff^T \bld{H}_{\ff\ff} \ff + \uu^T \bld{H}_{\uu\ff} \ff,
\end{aligned}
\end{equation}
where we truncate after the quadratic (bilinear) terms. Here, we use $\bld{g}_{x}$ to denote the gradient with respect
to the full cartesian nuclear coordinates (denoted by $\uu$), and $\bld{H}_{uu}$ for the full nuclear cartesian Hessian (denoted by $\uu$). 
We aim to predict the new equilibrium geometry in response to an applied electric field, i.e.
\begin{equation}
\nabla_{\uu}E(\uu,\ff) = \bld{0} = \bld{g}_{\uu} + \bld{H}_{\uu\uu} \uu + \bld{H}_{\uu\ff} \ff
\end{equation}
Assuming that the initial geometry has been optimized, the gradient $\bld{g}_{\uu}$ is zero and thus we can solve easily
for the displacement, 
\begin{equation}
\label{eq:xvec}
\uu = -{\bf{H_{\uu\uu}}}^{-1} \cdot \bf{H_{\uu\ff}}\cdot \ff
\end{equation}
This equation for \uu{} is quite useful, because differentiation with respect to \ff{} and $s_0$ (the length of the material
line) under the constraints of zero field and strain (which is consistent with the assumption that the energy is minimized) 
will yield the piezoelectric coefficient, as desired.  

Following through and taking the derivative with respect to \ff{} yields the matrix \firstDeriv{\uu}{\ff} 
\begin{equation}
  \label{eq:dxvec}
  \left( \firstDeriv{\uu}{\ff} \right)_{\bld{f} = \bld{0}, \bld{T} = \bld{0}} 
  = -\bld{H}_{\uu \uu}^{-1} \cdot \bld{H}_{\uu \ff}
\end{equation}
This equation can be seen as a generalization of eq. (6) from our previous publication \cite{Werling0} to the full dimensionality of the
potential energy surface.  
We note that even if the gradient were not ignored in 
\eq{eq:xvec}, it would disappear upon differentiation anyways. 
Furthermore, this equation, if evaluated at zero field is exact and
follows from the multivariable cyclic rule of calculus (the three vectors of interest are the gradient, the displacement, and the field), and 
this can be seen by keeping higher order terms and setting $\uu=0$ and $\ff=0$ in the result.

Although \eq{eq:dxvec} is formally correct, calculating $\firstDeriv{\uu}{\ff}$ in practice requires us to remove
translations and rotations from the coordinate system. 
Taking this step ensures (a) the physicality of the result, since
a molecular rotation or translation is typically made impossible by external mechanical constraints of the system, and
(b) that there is no problem due to singularity of the geometric Hessian $\bld{H}_{\uu{}\uu{}}$. 

At this point, it is useful to illustrate the physical picture underlying the derivation presented
so far. 
By rearranging \eq{eq:xvec}, we obtain 
\begin{equation}
  \label{eq:forces}
  \bld{H}_{\uu \uu} \uu = -\bld{H}_{\uu \ff} \ff
\end{equation}
The Hessian matrix contains the second derivatives of the energy with respect to nuclear positions, and hence, the term
on the left is the change in force (or gradient) generated by moving the
nuclei by the vector \uu from equilibrium (within the harmonic approximation).  
The term on the right is the approximate change in force generated by the field vector \ff on the nuclei.
At equilibrium (minimum energy), these forces must be equal. 
Since the geometric Hessian is a symmetric matrix, it has orthogonal eigenvectors, and 
one can represent the displacement vector \uu{} as a linear combination of these eigenvectors. 
Since translations and rotations correspond to zero or near-zero eigenvalues, respectively, one can construct infinitely
many displacement vectors that solve \eq{eq:forces} and only differ by the weight of the translation and rotation
eigenvectors. 
We also note that strain should not depend on translations and rotations (if Coriolis forces are
neglected). 
Aside from avoiding numerical problems in the inversion, it is therefore useful to remove translations and
rotations to obtain unique solutions for the displacement vectors. 
Since we are merely interested in describing strain, all that matters for the is the relative (intramolecular) deformation of the constituents of the system.

We will not discuss in detail how to construct translations and rotations to remove from the Hessian matrix, but we refer the reader to Appendix B.
We will now give a general outline of how to remove these vectors from the hessian.
Projecting out rotations and translations is easily accomplished by first constructing translation vectors and rotation vectors for the system. 
The Appendix gives a detailed account on how to construct rotation and translation vectors for the system in the nuclear cartesian space.
We use a tensor similar to the moment of inertia tensor (but instead mass independent)
its eigenvectors to construct the rotation vectors plus three translation vectors (altogether 5
vectors for linear and 6 vectors for non-linear systems). 
We may then choose a basis from which we project out these rotations and translations.  
The eigenvector of the Hessian are suitable although any basis which spans the full nuclear space will do.
The Appendix outlines how we then create a basis of either $3N-5$ (linear) or $3N-6$ (nonlinear) vectors which are orthogonal to our rotation and translation vectors.
For convenience we choose this basis to be orthonormal.  
We then organize this basis into the column vectors of a $n\times m$ matrix ${\bf{V}}$ ($n=3N$ and $m=3N-5$ or $m=3N-6$) .
It is important to note that this basis only lives in the vibrational space of the molecular system.  
Because the basis is orthonormal we have $\vv = {\bf{V}}^T\uu$. 
We then transform \eq{eq:dxvec} into the new coordinate system which occupies a subspace of the original system 
\begin{equation}
\begin{split}
\label{eq:HessTrans}
\bld{V}^{T} \cdot \bld{H}_{\uu \uu} \cdot \bld{V} \cdot \bld{V}^{T} \cdot \uu &= - \bld{V}^{T} \cdot \bld{H}_{\uu \ff} \cdot \ff \\
{\bf{H_{\vv \vv}}} \cdot \vv &= -{\bf{H_{\vv \ff}}} \cdot \ff
\end{split}
\end{equation}
The geometric Hessian on the left-hand side is now of reduced dimensionality $m \times m$, and $\bld{H}_{\vv \ff}$ is of
dimension $m \times 3$.  
We leave the coordinates for the field unchanged because ${\bf{H_{\vv\ff}}}$ is a two-point tensor, and we can choose the basis of \ff to be whatever we like.  
We usually choose them to be Cartesian coordinates which are in the same direction as the Cartesian coordinates of the nuclei for analysis purposes.

Now that ${\bf{H_{\vv\vv}}}$ is, in general, no longer singular, we can safely calculate the inverse (although this is
not necessary or the most efficient approach for solving this type of problem), to solve for the displacements in our new
coordinates, 
\begin{equation}
\label{eq:veq}
\vv = - \, \bld{H}_{\vv \vv}^{-1} \cdot {\bf{H_{\vv \ff}}} \cdot \ff
\end{equation}
There can be cases where the initial geometry optimization finds a saddle point instead of a minimum in the energy.  In
this case there may be additional modes which correspond to negative eigenvalues which may be projected out in addition
to rotations or translations. 

To convert the displacements back to Cartesian nuclear coordinates, one simply needs to multiply by $V$ from the left, 
\begin{equation}
\uu_{\mathrm{vib}} = \bld{V}\cdot \vv = - \, \bld{V} \cdot \bld{H}_{\vv \vv}^{-1} \cdot \bld{H}_{\vv \ff} \cdot \ff,
\end{equation}
where we introduce the subscript $\mathrm{vib}$ to signify that these displacements correspond only to intramolecular
deformations. 
Again, we differentiate with respect to the field under the constraint in the zero-field and zero-strain limit to obtain 
\begin{equation}
  \label{eq:finalDxdf}
  \left( \firstDeriv{\uu_{\mathrm{vib}}}{\ff} \right)_{\bld{f} = \bld{0}, \bld{T} = \bld{0}} = 
  - {\bf{V}} \cdot {\bf{H_{\vv \vv}}}^{-1} \cdot {\bf{H_{\vv \ff}}}
\end{equation}
\eq{eq:finalDxdf} is our final result, and we now have only to concern ourselves with retrieving the final
formulas for the  piezoelectric matrix for small systems.

\subsection{The ${\bf{P}}$ matrix for Molecular Systems}
We have thus far discussed the derivation of the Piezoelectric matrix ${\bf{P}}$ from a discussion of continuum mechanics, and how it encapsulates
the deformation properties along a material line at a given point in a continuum body. 
We have calculated $ \firstDeriv{\uu_{\mathrm{vib}}}{\ff}$ for a molecular system as a replacement to calculating the tensor field (can be specified for every point in a 
continuum body) $\frac{\partial \bld{u}}{\bld{f}}$ for a continuum body.
We have also discussed calculating the piezoelectric coefficient as the derivative of the longitudinal strain of a material line with respect to field, which we then evaluate
at zero field. 
Now that we have $ \firstDeriv{\uu_{\mathrm{vib}}}{\ff}$, we can present how to calculate a molecular version of $\bld(P)=\frac{\partial^2\bld{u}}{\partial s_0 \partial \bld{f}}$
to describe the field deformation characteristics around the molecular equivalent of a material line. 

To calculate a ${\bf{P}}$ matrix, we start by picking a material line in our system.  
This part can be (somewhat) tricky and has an incredible impact on the piezocoefficient.  
In a crystal it makes sense to study the
deformation of a unit cell due to the periodic nature of the crystal, since we expect that within a uniform field the deformation of all of the 
images of the unit cell should be identical.
There is no such natural kernel in the finite system realm.  
We have had much success in previous work in approximating the piezocoefficient by studying the deformation of the attribute of our
system we expect to have the most deformation within a field. 
One should take care however, in the assumption that the deformation properties of a small system extend to a bulk material.
However, if we are to compare similar molecular systems for their deformation properties,
it is reasonable to assume that picking a similar attribute in a group, such as the hydrogen-bond in our previous systems, would \textit{a priori}, be a good way to establish which members of
the group are the best piezoelectrics.

For our purposes here, we need only be concerned with a general method by which we choose a material line.  We shall, then with our freedom,
choose our material line to be a line segment between two atoms.  
We choose the two atoms via chemical intuition and the deformation we expect
(ie. hydrogen-bonded atoms and the like). 
One of the atoms will act as the point $P_0$ in \fig{fig:deform} in the undeformed system and will thus serve as the point of origin for our
material line.
We could also form linear combinations of points and for instance consider a line segment between the centers of mass for two monomers
in our system, but we need not discuss this at this point.  
Now that we have our line segment, we recall the knowledge we have accumulated about deformation
analysis to aid in our efforts. 
We wish to approximate $\bld{P}=\frac{\partial^2\bld{u}}{\partial s_0 \partial \bld{f}}$ around the point $P_0$ of our material line.
We have the $3N\times 3$ matrix $ \firstDeriv{\uu_{\mathrm{vib}}}{\ff}$ wich holds the $3\times3$ matrix $\firstDeriv{\uu}{\ff}_i$, indicating the 
displacement field derivative for the $3$ nuclear coordinates with respect to the field coordinates, for the $i$th atom of the molecule or system. 
We have dropped the ''vib'' subscript for convenience, though it is understood.
The molecular version of the ${\bf{P}}$ for this material line in the vicinity of these two atoms is then given by
\begin{equation}
\label{eq:INeedIt2}
 P = \frac{\firstDeriv{\uu}{\ff}_2 - \firstDeriv{\uu}{\ff}_1}{r_0}  
\end{equation}
where $r_0$ is the distance between these two atoms in the equilibrium geometry. 
This is equivalent to a numerical derivative, though we can not arbitrarily choose how small to make $r_0$, but are handcuffed by the distance in the 
equilibrium geometry (a consequence of the discrete nature of the system).  It should be pointed out that the difference of any linear combination
of matrices $\firstDeriv{\uu}{\ff}_i$ can be used to calculate a piezoelectric matrix.  
The matrix would then coincide with the deformation properties of a material line connecting the two points given by the same linear combinations of ${\bf{u}}_i$
for the molecular system.  
For example the geometric center or the center of mass of two regions of a bigger molecular system can be used and the deformation properties 
for the material line between these two points can be determined.

As discussed in the previous sections, if we wished to calculate \dtt in a method similar to our other papers, we must contract the ${\bf{P}}$ matrix 
over the indices for the displacement vector with unit vector ${\bf{e}} = \frac{{\bf{r}}_2- {\bf{r}}_1}{r_0}$.  
Furthermore, we would also need to contract over 
the field indices, multiplying by a unit vector also in the direction of the ${\bf{e}}$, (the basis chosen for the field must coincide with the basis for the 
nuclear coordinates if the contracted vector is actually ${\bf{e}}$).
We would thus obtain.
\begin{equation}
\begin{split}
 \dtt &= {\bf{e}}_u{\bf{P}}{\bf{e}}_f \\
 &=\frac{\partial^2 u_{r_0}}{\partial r_{0} \partial f_{r_0}}
\end{split}
\end{equation}
The subscripts attached to the vector ${\bf{e}}$ are to indicate both the indices of contracation (ie. the vector $\uu$ or $\ff$) and to indicate that 
although the vector ${\bf{e}}$ is in the direction of the material line it might have two different representations depending on the nuclear coordinate
and field bases.
As an example, the $3\time3$ piezoelectric matrix were calculated for two atoms along the z-axis, and the field basis was chosen to correspond with the x,
y, and z direction, then the ${\bf{e}}=(0,0,1)$ and the component of the matrix $P_33$ would be equal to $\dtt$ as in our previous papers.  
At this point we should again mention that two piezoelectric matrices for two material lines (line segments between atoms) for a molecule can be 
very different even if the material lines occupy a similar region of space and are in similar directions.
It is for this reason that there is no way to logically approximate the full third rank piezoelectric tensor for a molecule.  
In a sense, the derivation for the piezoelectric matrix and the connection to our calculated piezoelectric matrix is somewhat tenuous, but from 
a philosophical point of view and a practical point of view the authors of this paper believe that the connections hsould be made.  

To this end the piezoelectric matrix also has significant utility beyond approximations of $\dtt$.  
We can ask questions like, what direction of applied field is necessary to get the largest possible deformation betweent two atoms of a molecular
system?
If we apply a small field $\ff$ to our molecule, ${\bf{P}}\cdot \ff$ will tell us the relative direction of the displacemement vectors of the two nuclei which make up our material line per unit of the distance separating the two nuclei.
If we were to diagonalize ${\bf{P}}$, we would obtain eigenvectors which indicate in what direction to apply a field to have our strain vector $ \frac{\partial {\bf{u}}}{\partial r_0}$ to be in the same direction as the field.  
This is of course not the same as optimizing the deformation response for an applied field.  
To do so we can optimize $\lvert\frac{\partial {\bf{u}}}{\partial r_0}\rvert^2$ under the constraint of a small finite field--namely, ${\bf{f}}^T{\bf{f}}=k$, where k is some arbitrary small square
of the magnitude of the field.
We construct the equation to optimize with the appropriate lagrange equation and set the gradient with respect to the field coordinates equal to zero.  
The result 
is the typical eigenvalue problem for the matrix ${\bf{P}}^T{\bf{P}}$
\begin{equation}
 \begin{split}
 \lvert\frac{\partial {\bf{u}}}{\partial r_0}\rvert^2&={\bf{f}}^T{\bf{P}}^T{\bf{P}}{\bf{f}} \\
 g({\bf{f}}) = {\bf{f}}^T{\bf{f}}-k&=0\\
 \end{split}
\end{equation}
 The constraint equations is given by $g({\bf{f}})$ along with the equation we wish to optimize.
 We combine the two by multiplying $g({\bf{f}})$ by $\lambda$ (lagrange multiplier) and subtracting the two equations to obtain our lagrange equation.
 We then differentiate with respect to the field and set the result equal to zero.   
\begin{equation}
 \begin{split}
  L({\bf{f}})&={\bf{f}}^T{\bf{P}}^T{\bf{P}}{\bf{f}}-\lambda({\bf{f}}^T{\bf{f}}-k)\\
  \frac{\partial g({\bf{f}})}{{\bf{f}}}&=0 \implies
  {\bf{P}}^T{\bf{P}}{\bf{f}}=\lambda {\bf{f}}
 \end{split}
\end{equation}

The matrix, ${\bf{P}}^T{\bf{P}}$, is symmetric and real, and therefore has orthogonal eigenvectors and real eigenvalues.   
The largest eigenvalue gives the maximum value of $\lvert\frac{\partial {\bf{u}}}{\partial r_0}\rvert^2$ for an applied field of 
magnitude one unit and the eigenvector indicates the direction in which to apply the field to get the optimum deformation.  
Since we construct ${\bf{P}}$ for two atoms in the in the molecular or 
small system, it is possible to find the optimum field direction to apply to the system to get the best deformation for any pair of atoms.  
The direction of the relative change in the displacement vector
between the two atoms will have the direction given by ${\bf{P}}{\bf{f}}$ 
which is not necessarily in the direction of the distance vector between the two atoms. 

\subsection{Full Molecular Piezoelectric Response as a Rank 4 Tensor} 
As mentioned in previous sections it is impossible to record a full third rank piezoelectric tensor for a given point in space for a molecular 
system--in contrast to a continuum body where we can calculate the piezoelectric tensor at every point in the continuum body as a tensor field.
We can, however, write down a piezoelectric matrix for every pair of atoms in a molecular system and store the results as a rank 4 tensor.
To describe the piezoelectric response of the molecule in its entirety, we organize the individual response matrices
into a supermatrix $\bld{D}$ (or rank-4 tensor $D_{ijkl}$) 
\begin{equation} 
  \bld{D} \equiv 
            \left(
              \begin{array}{ccccc}
                \bld{P}_{1,1}   & \bld{P}_{1,2}   & \cdots & \bld{P}_{1,N-1}   & \bld{P}_{1,N}   \\
                \bld{P}_{2,1}   & \bld{P}_{2,2}   & \cdots & \bld{P}_{2,N-1}   & \bld{P}_{2,N}   \\
                \vdots          & \vdots          & \ddots & \vdots            & \vdots          \\
                \bld{P}_{N-1,1} & \bld{P}_{N-1,2} & \cdots & \bld{P}_{N-1,N-1} & \bld{P}_{N-1,N} \\
                \bld{P}_{N,1}   & \bld{P}_{N,2}   & \cdots & \bld{P}_{N,N-1}   & \bld{P}_{N,N}   \\
              \end{array}
            \right)
\end{equation}
where $N$ is the total number of atoms.
An entry $P_{ij}$ in the supermatrix is given by
\begin{equation}
 P_{ij} = \frac{\firstDeriv{\uu}{\ff}_j - \firstDeriv{\uu}{\ff}_i}{r_{ij}} 
\end{equation}
where $\firstDeriv{\uu}{\ff}_i$ is the portion of the field derivative of the displacement vectors for the system corresponding to atom $i$ (as discussed 
in the previous sections), and 
$r_{ij}$ is the distance between atom $i$ and $j$ in the equilibrium geometry.

The diagonal elements are undefined (dividing a zero matrix by zero), and flipping the
row and column indices inverts the order of the subtraction of $\firstDeriv{\uu}{\ff}_i$ and $\firstDeriv{\uu}{\ff}_j $ and hence the 
supermatrix is antisymmetric
\begin{equation} 
  \bld{D} \equiv 
            \left(
              \begin{array}{ccccc}
                  \bld{P}_{1,1}   &   \bld{P}_{1,2}   & \cdots &   \bld{P}_{1,N-1}   &   \bld{P}_{1,N}   \\
                 -\bld{P}_{1,2}   &   \bld{P}_{2,2}   & \cdots &   \bld{P}_{2,N-1}   &   \bld{P}_{2,N}   \\
                  \vdots          &   \vdots          & \ddots &   \vdots            &   \vdots          \\
                 -\bld{P}_{1,N-1} &  -\bld{P}_{2,N-1} & \cdots &   \bld{P}_{N-1,N-1} &   \bld{P}_{N-1,N} \\
                 -\bld{P}_{1,N}   &  -\bld{P}_{2,N}   & \cdots &  -\bld{P}_{N-1,N}   &   \bld{P}_{N,N}   \\
              \end{array}
            \right)
\end{equation}
with $N(N+1)/2$ independent elements.
A similar analysis can be used to determine the field direction which optimizes the deformation between the two atoms for each ${\bf{P}}$ matrix.
Such a supermatrix can be useful for identifying which pairs of atoms could be the most useful in identifying ``good '' piezoelectric candidates
for similar families of systems.  

\section{Outline of Computational Procedure}

\begin{figure*}
  \centering

  \includegraphics[width=0.70\textwidth]{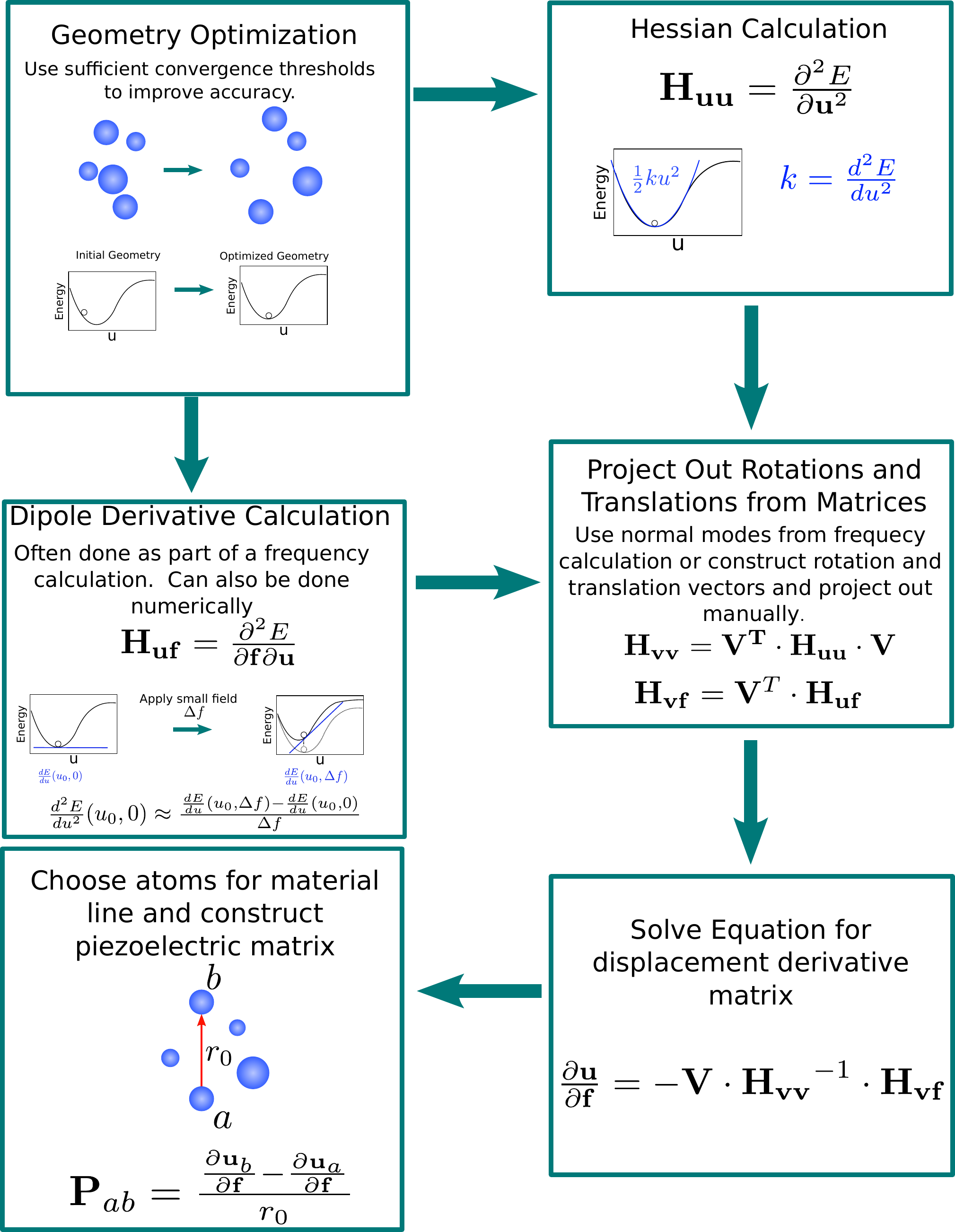}
  \caption{flowchart for computational procedure }
  \label{fig:flowchart}
\end{figure*}

(flowchart)
The procedure for calculating the piezoelectric matrix (consider renaming) will follow
that presented in \fig{fig:flowchart}.

\subsection{Geometry Optimization}

The first step is of course to optimize the geometry of the molecule/s or small system.  
Depending the the degree of accuracy desired for the calculation, it may be necessary to use
more stringent thresholds for gradient of the energy and especially displacement (since we are mainly
interested in approximating displacement under the application of an applied field).  
Furthermore, if one is performing a density functional theory (DFT) calculation for the energy, it might be necessary
to use a finer grid for the integration and smaller cutoffs for the integrals. 
We have noticed previously that without a decent grid and low cutoffs that there tends to be small ``bumps'' in  cross-sections of the 
energy which may interfere with calcuations of the hessian and other derivatives with respect to displacement.

Furthermore, one might be interested in constraining the system in some position which might be relevant to the crystal 
structure or some other reason.  
For these purposes it might be necessary to employ different methods for constraining the
geometry and still performing the geometry optimization.  
These might include lagrange multiplier or projection techniques.
For these we prefer the reader to \cite{numRec, schlegel, baker}.  
It is important to remember though that in later steps one might want to 
later project out such unwanted degrees of freedom in addition to rotations and translations.
Any type of motion corresponding to an unwanted deformation can be constructed just like the rotation and translation vectors in the Appendix and 
also projected out of the basis.  
The corresponding transformation column vector matrix ${\bf{V}}$ will then be reduced in columns by the number of additional unwanted vectors.  
This would of course not hold if these other unwanted motions are linearly dependent with the rotations and tanslation vectors.
\subsection{Hessian Calculation}

Frequency calcualtions have become statndard in many program packages and often the Hessian matrix is reported at the 
end of the calculation or can be recovered from scratch files.  
Analytical Hessian calculations
might be prohibitively expensive for large systems and parallel implementaions are not always present for some methods in software packages.
However, numerical calculations of the Hessian are always 
embarassingly parallel due to the vast number of independent gradient calculations.  
We have found for our calculations that molecules greater than 25 atoms or so start to become
prohibitively expensive for analytical B3LYP \cite{b3lyp,Lee1988} calculations with a moderately
sized basis set (6-31G(d))\cite{pople}.
Furthermore, we do not notice a remarkable discrepancy in the 
final calculation of the piezoelectric matrix when using a numerical Hessian.  

\subsection{Dipole Derivative Calculation}

The dipole derivative, $\frac{\partial^2 E}{\partial \uu \partial \ff}$, calculation may be performed in parallel with the Hessian calculation.
Alternatively, if one is performing a vibrational analysis for the Hessian calculation, the 
dipole derivatives are usually calculated as well since they have a well established 
relationship with the intensity of infrared active vibrational modes \cite{kom}.  
If this is the case, it is likely that one can merely retrieve the dipole derivatives with the Hessian 
at the end of a frequency calculation. 

If one wishes to perform the dipole derivative calculation manually we refer the reader
to Ref \cite{kom}, and we also present the following procedure.  
After the geometry optimization, either retrieve or perform a separate calculation to obtain the gradient ($\nabla E(\ff=0)$) of the energy. 
After this, one can calculate the gradient of the molecule for three separate field calculations.  
The most reasonable choices for field directions are the x, y and z direction.  
The field magnitude, which we call here $df$,  applied in each direction should be quite
small (~0.001V/nm).  
After the three calculations are finished, one may approximate $\frac{\partial \nabla E}{\partial f_x}$,$\frac{\partial \nabla E}{\partial f_y}$ and 
$\frac{\partial \nabla E}{\partial f_z}$, (or the corresponding length variables for whatever basis you choose) using the finite differences of the gradient from zero field.
\begin{equation} 
\begin{split}
\frac{\partial \nabla E}{\partial f_x} &\approx \frac{\nabla E(f_x = df, f_y=0, f_z=0)-\nabla E(\ff=0)}{df} \\
\frac{\partial \nabla E}{\partial f_y} &\approx \frac{\nabla E(f_x = 0, f_y=df, f_z=0)-\nabla E(\ff=0)}{df} \\
\frac{\partial \nabla E}{\partial f_z} &\approx \frac{\nabla E(f_x = 0, f_y=0, f_z=df)-\nabla E(\ff=0)}{df}
\end{split}
\end{equation}

Each of these derivatives is a vector quantitiy of dimension $3N$, where $N$ is the number of nuclei, and represent the columns of the matrix ${\bf{H_uf}}$ as presented in the previous sections 
and (flowchart figure).
Alternatively, one might choose to perform a few gradient calculations at a few different small field strengths for each field direction and perform a linear regression on each component of the gradient
to find the slope of each component. 
We did not notice that this drastically improves the result and is usually unnecessary.  (consider adding the graphs for this)

\subsection{Project Out Rotations and Translations}

Now that we have the Hessian ${\bf{H_{uu}}}$ and the Dipole derivative ${\bf{H_{uf}}}$ matrices, we need to project out rotations, translations and any other unwanted modes
(if using constraints)from the matrices.  
This is equivalent to transforming the the basis of the displacement variables to a subspace of modes which lack displacement corresponding to rotation
and translation.  
A typical unitary transformation of the Hessian for example (${\bf{H_{UU}}}={\bf{U}}^T \cdot {\bf{H_{uu}}} \cdot {\bf{U}}$) would transform the Hessian from derivatives
$ H_{u_iu_j} = \frac{\partial^2 E}{\partial u_i \partial u=j}$ to $H_{U_iU_j}=\frac{\partial^2E}{\partial U_i \partial U_j}$, where the variables of the transformed matrix are now with respect
to the length variables of the new basis.  
Here we consider that ${\bf{U}}$ is unitary and made up of orthonormal column vectors.  
In the case where we ignore rotations and translations, we must construct the matrix ${\bf{V}}$ discussed in previous sections by projecting out translation and rotation
vectors from the eigenvectors of the the Hessian (or any other basis that spans the full nuclear space).  

  In practice, if one is using modern quantum chemistry software, it might be possible to obtain cartesian normal modes from job outputs or the scratch directory.  
 If this is the case the vectors may already be sorted so that the vectors corresponding to rotations and translations are reported first or last.  
 If this is not the case, one may always apply the Hessian to 
the normal modes, and record the norm of the results.  
The vectors corresponding to the smallest norms (ie. very close to zero) are typically the vectors which represent translations and rotations.
If a molecule is linear we expect there to be 5 such modes, and ${\bf{V}}$ will have dimesions $3N\times3N-5$.  
If the molecule is nonlinear we expect 6, and ${\bf{V}}$ will have dimesions $3N\times3N-6$.
If the translational modes and rotational modes have been identified in this way, the remaining normal modes are make up the columns of ${\bf{V}}$. 
One should also check that these vectors are orthonormal.  
Using ${\bf{V}}$, we may now transform ${\bf{H_{uu}}}$ and ${\bf{H_{uf}}}$ to ${\bf{H_{vv}}}$ and ${\bf{H_{vf}}}$ according to (flowchart and equation).

If the program package one is using does not offer cartesian normal modes after a frequency calculation, it is possible to construct translational and rotational modes which may then 
be projected out from the eigenvectors of the Hessian or mass weighted Hessian.  
This should then lead to $3N-5$ or $3N-6$ (for linear or nonlinear molecules respectively) nonzero 
vectors which may then be internally orthonormalized via the Gram-Schmidt method or otherwise \cite{numRec}.
It might be the case that after projecting out the rotations and translations that there are not $5$ or $6$ zero vectors.
After the orthogonalization routine, there should be this many zero vectors.  
For more information, we refer the reader to Ref \cite{numRec}.

\subsection{Solve for Displacement Derivative Matrix}

Now that we have the transformed Hessian and dipole derivative matrix (${\bf{H_vv}}$ and ${\bf{H_vf}}$ respectively), we may now solve for the displacement derivative matrix 
$\firstDeriv{\uu_{\mathrm{vib}}}{\ff}$.  
This matrix contains the derivatives of all the nuclear displacement vectors ${\bf{u}}$ with respect to the field vector ${\bf{f}}$ under condition
that the energy remains minimized and only deformations in the space of the vibrational motion of the molecule is allowed.  
The matrix is inherently $3Nx3$ once put back into normal cartesian space.  
We generally solve in the vibrational mode space, but simultaneously 
transform back to the cartesian space.  
The equation is given by \eq{eq:finalDxdf} and is also included in \fig{fig:flowchart}.  
Even though the inverse of the ${\bf{H_vv}}$ is written in \eq{eq:finalDxdf}, 
it is not necessary to invert this matrix; it is generally more efficient to just solve equation (equation) for $\frac{\partial {\bf{v}}}{\partial {\bf{f}}}$ via known algorithms, widely implemented in such programs as Mathematica, Matlab, and Python
packages \cite{math,MATLAB:2010,python}.  
Then of course one must transform $\frac{\partial {\bf{v}}}{\partial {\bf{f}}}$ back to normal space via ${\bf{V}}\cdot \frac{\partial {\bf{v}}}{\partial {\bf{f}}}$.

\subsection{Construct Piezoelectric Matrix}

As discussed in the previous Derivation section, to construct the final piezoelectric matrix ${\bf{P}}$, one first needs to define a ``material line'', 
which in the molecular sense we 
take to mean the line segment connecting two atoms of choice in the system (or some other points of reference which will change upon deformation).  
The unit vector in our position coordinate
system then becomes ${\bf{e}}$.  
From the matrix ${\bf{P}}$ we can describe the relative deformation of these two points.  
${\bf{P}}$, to reiterate, will be an approximation to
$\frac{\partial^2 {\bf{u}}}{\partial r_0 \partial {\bf{f}}}$, which we approximate via the change in the derivative of the displacement 
vector ${\bf{u}}$ with respect to the field vector
${\bf{f}}$ as we move from atom ``a'' to atom ``b'', per unit distance as we move from one atom to another along the ``material line'' in the undeformed body.
The calculation is simple and is given by
\eq{eq:INeedIt2} and \fig{fig:flowchart}.  
The only difficult task is identifying which parts of the $3Nx3$ displacement derivative matrix $\frac{\partial {\bf{u}}}{\partial {\bf{f}}}$ correspond to 
$\frac{\partial {\bf{u}}_a}{\partial {\bf{f}}}$ and $\frac{\partial {\bf{u}}_b}{\partial {\bf{f}}}$.  
Both parts are $3x3$ and hence make ${\bf{P}}$ also $3x3$.  

Finding these parts, however, depends on the order of position variables used by the software to write the Hessian and gradients of the molecule.  
Typically the ordering is 
(atom1 x, atom1 y, atom1 z  . . . atomN x, atomN y, atom N z). 
So if one uses for example atom 3 for ``a'' and atom 35 for ``b'' as the two ends of the ``material line'',  $\frac{\partial {\bf{u}}_a}{\partial {\bf{f}}}$
will be given by rows $3*2+1=7$ through $3*2+3=9$, and $\frac{\partial {\bf{u}}_b}{\partial {\bf{f}}}$ will be given by $3*34+1=103$ through $3*34+3=105$ of $\frac{\partial {\bf{u}}}{\partial {\bf{f}}}$.  
Hence, $\frac{\partial {\bf{u}}_i}{\partial {\bf{f}}}$
corresponds to rows $3*(i-1)+1=3i-2$ through $3*(i-1)+3=3i$. 
This of course assumes that we index the rows starting at $1$ and up to $3N$.  
In python and other languages it is common
to index starting from $0$ to $3N-1$. 
If this is the case, then atom 3 would actually be atom 2 (atom 0 and atom 1 come before) and atom 35 would be atom 34.  
In this instance the 
rows to extract from $\frac{\partial {\bf{u}}}{\partial {\bf{f}}}$ would be $3i$ to $3i+2$.

\section{Results and Discussion}

\begin{figure*}
  \centering

  \includegraphics[width=1.00\textwidth]{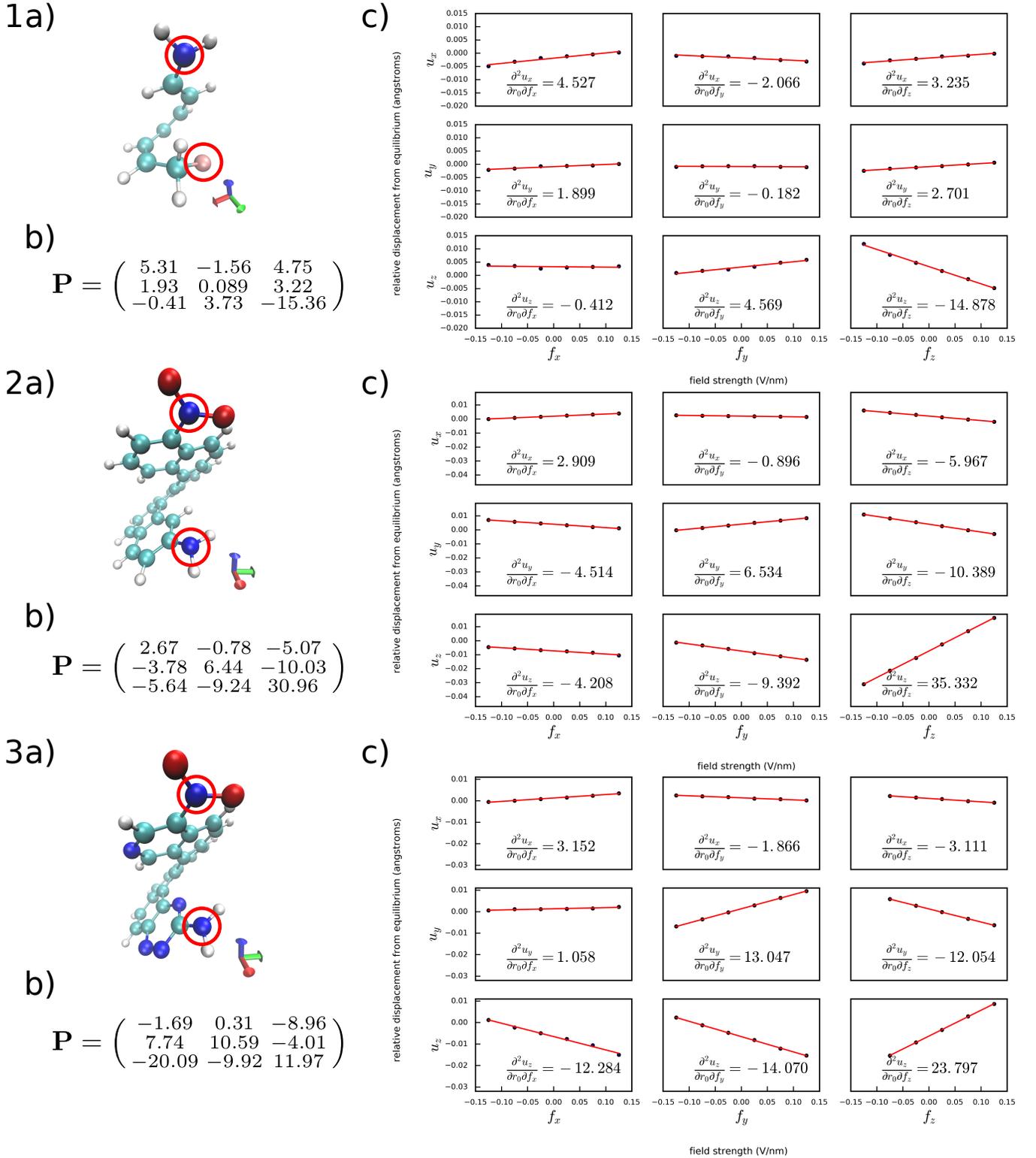}
  \caption{1 a) A small helicene-like structure is depicted with the two atoms chosen for the ``material line'' highlighted. b) The calculated piezoelectric matrix is given under the molecule. c) 
  An estimated matrix from geometry optimizations is shown for comparison on the right. 2 a) A traditional helicene molecule is depicted with the two atoms chosen for the ``material line''
  highlighted. b) The calculated piezoelectric matrix is given. c) An estimated matrix from geometry optimizations is shown for comparison on the right. 3 a) A nitrogen-rich helicene molecule is depected with the two
  atoms chosen for the material line highlighted.  b) The calculated piezoelectric matrix is given.  c) An estimated matrix from geometry optimizations is shown for comparison on the right. The axes are presented with 
  the molecules for an idea of how the two highlighted atoms move in a field as described by their matrices. All values for matrix components are given in $\frac{pm}{V}$. Molecules are rendered with Tachyon in VMD\cite{HUMP96, STON1998}.}
  \label{fig:matrix}
\end{figure*}
\begin{figure}
  \centering

  \includegraphics[width=0.48\textwidth]{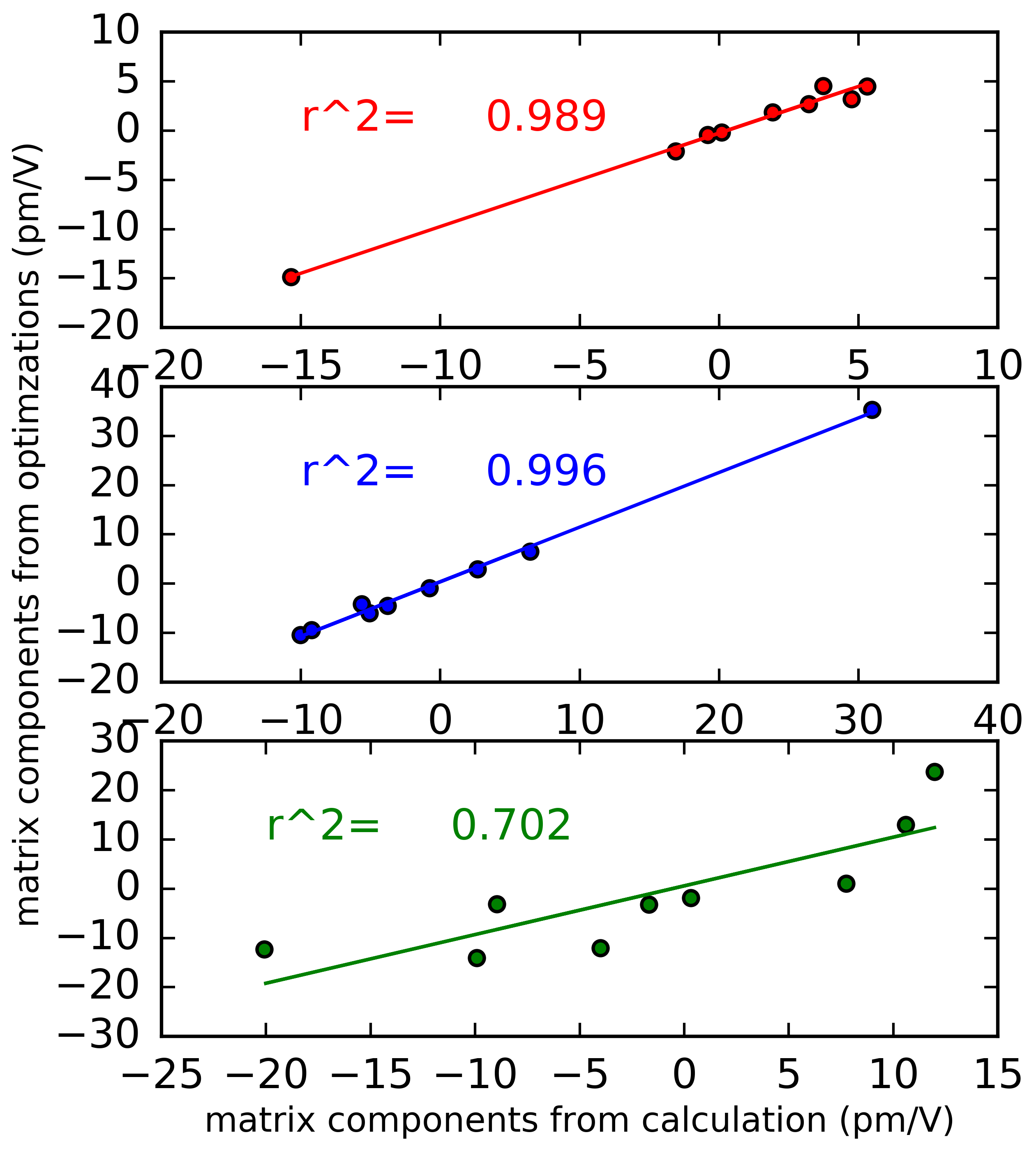}
  \caption{Linear regressions were performed between the matrix values shown in \fig{fig:matrix} between the matrices calculation with \eq{eq:INeedIt2} and the approximated values from the geometry optimizations.
  The $r^2$ values are given.  The red, blue and green plots correspond to the first, second and third molecules respectively as shown in \fig{fig:matrix}.}
  \label{fig:correlation}
\end{figure}

Three molecules were used to test the method developed to calculate a piezoelectric matrix ${\bf{P}}$. 
\fig{fig:matrix} shows the three molecules and the pairs of atoms in each which were used 
to calculate a piezoelectric matrix (a).  
The Piezoelectric matrix based on the atom pairs is also reported from our calculations outlined in the Procedure section (b).  
Furthermore,
we also report an estimated piezoelectric matrix from optimizations(c).  
All energy calculations were performed at the B3LYP/6-31G(d) level with QChem \cite{Qchem}.  
The plots shown in (c) correspond to geometry optimizations performed with fields applied in the x, y, and z direction
(columns).  
The rows of the matrix correspond to $u_x$, $u_y$, and $u_z$.  
A single plot shows the change in displacement ($u_{ai} -u_{bi}$ (atoms a and b  and component of displacement vector $u_i$))
from equilibrium between the two highlighted atoms for the corresponding 
coordinate $u_i$ (given by row) for varying values of field the corresponding field direction $f_j$ (given by column).  
All other field components are held at 0 for a given plot. 
The number reported in each plot is the slope of the plot divided by the equilibrium distance ($r_0$) for the two atoms; 
the numbers should be close to the corresponding entry in the 
reported ${\bf{P}}$ matrix.  
The molecules chosen are helicene-like because these molecules offer us the ability to test our new method on single molecule systems which were outside the abilities of our previous methods
The reason helicene molecules are interesting is 
because the interactions of the polar ends of the molecules are balanced by dispersion between the coils.  
Previous work by Huchison et al. \cite{quan}
shows that these molecules show considerable piezoelectric response.  
%
Furthermore the equilibrium (zero field) structure for each molecule was 
aligned with the z-axis such that the two atoms of interest lie on the z-axis.  
The chosen atoms were generally chosen to be part of the polar functional groups.  
We generally expected the response to be greatest for the displacment in the 
z direction with an applied z field (${\bf{P}}_{33}$) which was the case for the first two molecules. 
The last molecule, however, showed a large ${\bf{P}}_{31}$ component (3rd row and 1st column),
or in other words a substantial change in displacement in the z direction between the two atoms when a field was applied in the x direction.  
\fig{fig:correlation} shows the correlation between the calculated piezoelectric components and the components predicted by geometry optimizations.
The $r^2$ values are quite good for the first two molecules ($~0.99$), but quite poor for the last molecule ($~0.70$) (\fig{fig:correlation}).
Generally, the errors are quite small for the first two molecules, but the last nitrogen rich helicene shows significant errors in a number of matrix components.
The signs for the matrix components are generally in agreement except for some cases.  
It should be noted that when the piezoelectric matrix is computed for the third molecule at the Hartree-Fock/sto-3g \cite{fock, hehre, collins} level and the geometry optimizations performed, the 
agreement is very similar to those of the previous molecules ($r^2\approx 0.99)$.  
This might indicate that there are issues with performing density functional theory calculations at finite fields.
More specifically, a grid is used to calculate the exchange-correlation functionals in many cases and it might be possible that the electron density collects in a a region where electrons would no longer be considered bound.
In this case one expects that the grid extends past the relative maxima outside where the colombic potential of the nuclei approaches zero after which the energy due to the field goes to negative infinity.  
As mentioned before, molecules are not stable in electric fields and ionize in the limit of infinite time \cite{Yamabe1977}.
This problem is avoided for small systems by using local basis sets.  
Furthermore, we have noted that as basis set size increases, the correlation between the calculated piezoelectric matrix and the approximated matrix from geometry optimizations seems to 
decrease which might indicate that the calculation of the electron density within a field might become less and less accurate as the electrons are given more ``space'' to occupy. 
This conclusion seems dubius, however, because it is easily shown that the relative maxima for the nuclear potential within a field of 1V/nm for a system like hydrogen fluoride is on the order of 70 bohr away from the 
nuclei of the system, and the finite field values we use are typically much less than this.  
Therefore in the regin around the molecule where the grid is used for quadrature should appear only as a binding potential for the electrons.  
More work is needed to assess whether nonbinding affects effect these finite fiel geometry opitimizations.

There are numerous other possible sources of errors, but they are mainly retained by the matrix estimated from optimizations and not the method introduced in the previous sections. 
For example,
geometry optimizations are never ``fully'' optimized. 
Instead, the user usually specifies different tolerances for convergence: like a threshold for the gradient of the molecule, a threshold for the change in energy, or
a threshold for some metric which describes displacement of the molecule from a previous optimization step. 
When a few or all of requirements for optimization are met the molecule is considered to was noted that at relatively low fields in the 
be optimized.  
Often freqency calculations are performed to make sure that no imaginary frequencies are present (ie. there are no negative eigenvalues of the Hessian evaluated at the 
current geometry) which would indicate that the optimization has failed to produce a nuclear geometry which resides in a basin of the electronic potential energy surface.  
Using different 
convergence criterion, we found that the calculated piezoelectric matrices (b in \fig{fig:matrix}) showed very little fluctuation in components.  
The estimated matrices shown, however, tended to fluctuate
more radically (some values changed as much as 100 percent or more).  
Also, for DFT calculations, a fine grid should be used to evaluate the energy.  
In previous work \cite{Werling0}, we found that the potential
energy surface shows noticeable bumps when scanning across it (via moving nuclei), and this problem was resolved by using a finer grid to evaluate the density functional.  
This is especially
important if numerical evaluations of the Hessian are to be performed.  

Furthermore, there is one more perhaps surprising possible source of error in our calculations.  
If one looks closely at \fig{fig:matrix}, one can see that the linear regression plots
routinely do not cross $(0,0)$.  
However, the y-axis corresponds to the difference of the displacement coordinate $u_i$ between the two atoms of the ``material line'' compared to the 
undeformed system.  
This should mean that there is necessarily no difference in the displacement coordinates between the system at zero field and the undeformed geometry (which was optimized at zero field).  
The point $(0,0)$ is actually omitted from every plot because it does not fit with the rest of the data.  
However, the rest of the data is shifted in the same direction from this point, and the 
error is systematic.  
We have one possible theory for why this might be.  
If one aligns the molecule hydrogen fluoride in QChem along the axis, for example, and then apply a field in the z-direction, one expects
the molecule to align itself in the z-direction to minimize energy.  
QChem, however, removes translations and rotations from steps in the optimization, and for a two nuclei system, this 
does not allow the molecule to reorient itself within the field.  
If a large enough field is applied, however, it is possible that QChem will report the optimziation as never converged, despite the 
fact that the atoms themselves are no longer moving in the optimzation.  
This indicates that it is likely that QChem does not remove rotations from the gradient when using it for 
evaluation as to whether the geometry has converged or not.  
However, the optimization will seek to diminish the gradient nonetheless.  
To do this it must diminish the gradient along modes
other than rotations.  
This is possible because the gradient was non zero to begin with (the gradient never reaches zero for molecules in an optimization due to the time it would take).  
This means to converge the 
geometry, the molecule must move relative not just to an applied field but further along the unperturbed potential (if you view the new potential as a linear combination of the 
zero field potential and a new component due to the field). 
These movements would be systematically in the same direction because they correspond to movements along the unperturbed potential, and therefore the 
points in \fig{fig:matrix} would likely shift in the same direction despite the field direction.  
This is conjecture, but it might explain this shift off of $(0,0)$.  
For larger field magnitudes, 
it might also cause bigger shifts and might likely increase the resulting matrix components. 
With all of these things considered, the new method has none of these downfalls and is only 
limited by the time it takes to perform the Hessian calculation, which can be improved by using numerical methods present in most software over large numbers of cores.

There is one more system-specific reason why the calculated and esitmated piezoelectric matrices might disagree for the third molecule. 
It should be noted that for the optimzations performed for fields in the z-direction for the third molecule in \fig{fig:matrix} that the last point is omitted for the smallest
field value.  
This was because the hydrogens on the amine group flipped orientation.  
If we look at the steps of the geometry optimization, we can see that on of the modes of the approximated Hessian in the optimization routine, becomes negative at some point and then becomes 
positive after a few steps.  
This typically indicates that the molecule is transitioning between relative minima in the electronic potential.  
In this instance it is readily viewed as a flip in the amine hydrogen orientation. 
Since we omitted this point, this could not be what causes a large deviation in the $P_33$ value for this curve.  
However, it was noted that at some point in the optimization before the amine flip that the number of modes for the Hessian was decreased by 1 for a step.  
usually translations and rotations corresponding to 6 modes are left out, but in this case an extra mode was left out for a step. 
This should indicate that a mode was very close to having an eigenvalue near zero.  
Indeed two such modes are seen throughout the optimzation.  
The mode would be omitted if it is very close to zero because, when the hessian is inverted to form the next step it would lead to very large displacements corresponding to the mode.
This might indicate that this molecule might have one or two very close energy minima in the space of the nuclear displacements and at even small fields the molecule might be able to switch basins. 
This essentially changes the hessian and would lead to a deformation response from geometry optimizations different than that calculated from \eq{eq:finalDxdf}.
This would be very interesting for the deformation properties of this molecule in real experimental settings.
The zero field calculation of the piezoelectric matrix can not show such a phenomenon.
This theory is also conjecture and needs to be explored further, but it could indicate that smaller fields might be able to completely change the deformation properties of a molecular system.

\section{Conclusion and Outlook}
Up to this point, we have sought to explore the piezoelectric properties of organic piezoelectrics--which have promising applications in material science.
In two previous bodies of work we have shown that we can predict to well within an order of magnitude the piezoelectric properties of organic crystals like MNA
via methods that involve finite field optimizations and/or simple energy ``scans`` along a coordinate of interest (eg. hydrogen bonds).
While these methods are useful for monomeric systems in which one can use chemical intuition to select a coordinate of interest for which to measure the field dependent 
deformation properties, for single molecules and slightly larger systems, it may be difficult to identify and exploit such properties as intermolecular bonds which
otherwise leads us to calculate piezoelectric properties by doing time consuming geometry operations for large numbers of field magnitudes as well as directions.

In this work we have taken the mathematical model and equations resulting from it from our previous work and extended it to the full nuclear dimensionality of 
a small system.  
In this way, with just one geometry optimization and one frequency calculation (sometimes threee gradient calculations as well), we can aquire all of the necessary information
to describe how the molecule or system will deform in a small field and hence obtain the full piezoelectric properties for the system around zero field.
To this extent we do not need to use any apriori knowledge or intuition for a system and do not have to reduce the dimensionality of the system to understand its
field deformation properties.
Furthermore we have made deep connections to strain theory and have addressed the difficulty (or impossibility) of defining a full 3rd rank piezoelectric tensor for a small
system.  
To this end we show that we can approximate a piezoelectric matrix which acts like a contraction of the piezoelectric tensor with a unit vector in the direction of the vector between 
two atoms in the system, but due to the discrete nature of the system a full piezoelectric tensor field cannot be calculated, but instead we can calculate a piezoelectric matrix for each pair of atoms.
In a sense, the piezoelectric properties of a molecule or small system show extreme anisotropy and discontinuity.

However, we have seen that this method can be used to select for ''good`` organic piezoelectric candidates by screening similar families of molecules for specific deformation properties
of one or many different pairs of atoms (or linear combinations of atoms). 
To this extent we hope that this work serves as a staging point for further investigation into other areas of molecular or finite system piezoelectrics--like controlling oscilations in a field
or optimizing the work done by actuators, etc.  
Also we hope that the connections we have made to continuum strain theory will raise and may have already answered philosophical questions such as at what length scale can something be called 
piezoelectric and how do  and can we quantify properties like the piezoelectric tensor for such systems.

\begin{appendix}

\section{Relevant Strain Theory}
\begin{figure}
  \centering
  \includegraphics[width=0.48\textwidth]{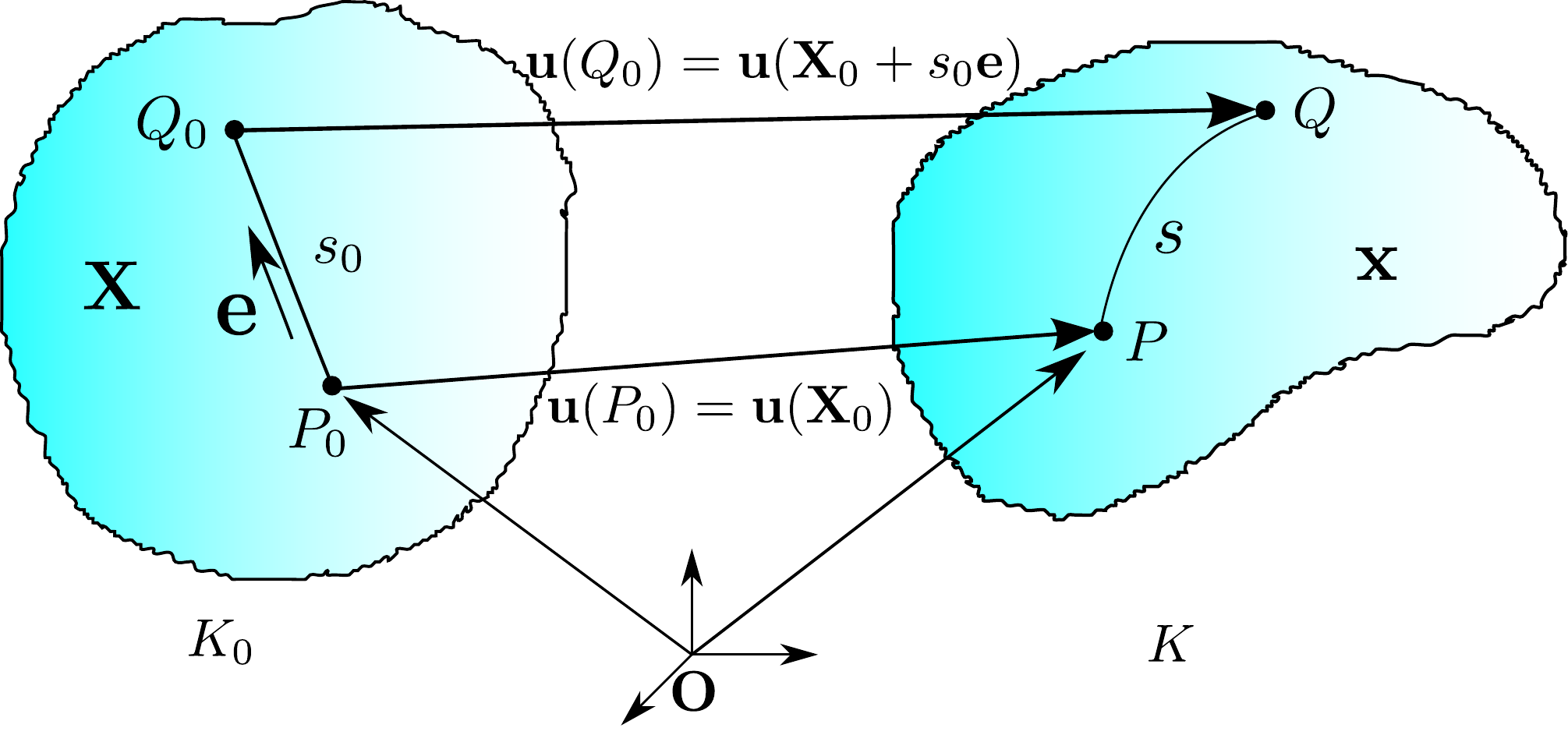}
  \caption{General deformation of a body.~\cite{contMech}}
  \label{fig:deform1}
\end{figure}
For convenience we again include ~\fig{fig:deform1}. 
Again, the material covered here closely follows that presented in Ref \cite{contMech}
Recall that $K_0$ and $K$ correspond to the undeformed and deformed body respectively, and that the line segment from 
$P_0$ to $Q_0$ in the undeformed body represents a material, which in the deformed body is given by the curve $P$ to $Q$.
Here we will derive the deformation gradient ${\bf{F}}$ (\eq{eq:dGrad}) and the Green strain tensor (\eq{eq:gst}) rigorously,
and rederive the equation for the longitudinal strain (\eq{eq:longstrain2}).

Recall that we wish to derive the Green Strain Tensor ${\bf{E}}$ and ~\eq{eq:dsds022}, which we later use to 
derive the final formula (\eq{eq:longstrain2}) for the longitudinal strain at a point $P$ of our body.
We will begin where we 
left off in the Background section (\eq{eq:dsds0}).  
We can examine how the components of a vector along $P_0Q_0$ change as we vary $s_0$, the length parameter of our undeforme material line.  
If we write this vector as ${\bf{r}}_0= X_k{\bf{e}}_k$, where ${\bf{e}}_k$ are the 
unit vectors which make up the basis, we have for a small change in the vector
\begin{equation}
\label{eq:dXkds0}
  \begin{split}
    d{\bf{r}}_0 &= {\bf{e}}\,ds_0=dX_k\,{\bf{e}}_k \\ 
    \Rightarrow \quad dX_k &= e_k\,ds_0 \\ 
    \Leftrightarrow \quad e_k &= \frac{dX_k}{ds_0}
  \end{split}
\end{equation}
We will use \eq{eq:dXkds0} later.

If we consider the vector differential along $s$, tangent to the curve, which we call $d{\bf{r}}$ and has components $dx_i$, we may write
\begin{equation}
\label{eq:drds0}
d{\bf{r}}=\frac{\partial{\bf{r}}}{\partial s_0} \quad \Rightarrow \quad dx_i=\frac{x_i}{\partial{s_0}}ds_0
\end{equation}
Now we would like to relate the line elements $d{\bf{r_0}}$ from $K_0$ with $d{\bf{r}}$ in $K$.
\begin{equation}
\label{eq:drd0}
d{\bf{r}}=\frac{\partial{\bf{r}}}{\partial {\bf{r}_0}} \cdot d{\bf{r}_0} \quad  \Leftrightarrow \quad dx_i=\frac{\partial x_i}{\partial X_k}dX_k
\end{equation}

We thus have the deformation gradient ${\bf{F}}$.
\begin{equation}
 \label{eq:defGrad}
 {\bf{F}}=\bigtriangledown({\bf{r}})=\frac{\partial{\bf{r}}}{\partial{\bf{r}_0}} \quad \Leftrightarrow  \quad F_{ik}=\frac{\partial x_i}{\partial X_k}
\end{equation}

We would like to calculate $(\frac{ds}{ds_0})^2$ from before.  
Noting that 
$ 
 \frac{\partial x_i}{\partial s_0}\bigg|_{s_0=0}
 =\frac{\partial x_i(X,t)}{\partial X_k}
 =\frac{\partial (X_k+s_0e_k)}{\partial s_0}\bigg|_{s_0=0}
$ leads to 
\begin{equation}
 \frac{\partial x_i}{s_0}=
 \frac{\partial x_i}{\partial X_k}\frac{dX_k}{ds_0}=F_{ik}e_k
\end{equation}
In the last step we have used our result from \eq{eq:dXkds0}.  
This is analogous to a directional derivative, but instead of using
the gradient of a scalar quantity, we instead use the the gradient of a vector quantity which is inherently a matrix--the deformation gradient.

We can now write
\begin{equation}
\label{eq:dsds02}
 \begin{split}
 \bigg(\frac{ds}{ds_0}\bigg)^2 &= \frac{dx_i}{d\bar{s}_0}\frac{dx_i}{d\bar{s}_0}  \\
 &= (F_{ik}e_k)(F_{il}e_l) \\
 &= {\bf{e\cdot (F^TF)\cdot e}} \\ 
 &= \bf{e\cdot C \cdot e},
 \end{split}
\end{equation}
where ${\bf{C}}$ is known as the Green deformation tensor. 
Recall
that the displacement of a particle during deformation is given by $\bf{u}(\bf{r}_0, t)$.
We would like to recast the deformation tensor in terms of the displacement gradient, which
we call $\bf{H}$
\begin{equation}
\label{eq:H}
 \bf{H} = \frac{\partial {\bf{u}}}{\partial {\bf{r}_0}} \quad \Leftrightarrow \quad H_{ik}=\frac{\partial u_i}{\partial X_k}
\end{equation}
Recalling that
\begin{equation}
 x_i(X,t) = X_i +u_i(X,t)
\end{equation}
we may infer
\begin{equation}
 \frac{\partial x_i}{\partial X_k} = \frac{\partial X_i}{\partial X_k}+ \frac{\partial u_i}{\partial X_k}
\end{equation}
or
\begin{equation}
 F_{ik} = \delta_{ik}+ H_{ik} \quad \Leftrightarrow \quad \bf{F} = \bf{1} + \bf{H}
\end{equation}
We recast the Green deformation tensor as follows:
\begin{equation}
 {\bf{C}}= {\bf{F^TF}} = ({\bf{1}} + {\bf{H^T}})({\bf{1}}+ {\bf{H}})= {\bf{1}}+ {\bf{H}} + {\bf{H^T}} +{\bf{H^TH}}
\end{equation}
The Green strain tensor, ${\bf{E}}$ is defined by:
\begin{equation}
\label{eq:H2E}
 {\bf{E}}= \frac{1}{2}({\bf{H}} + {\bf{H^T}} + {\bf{H^TH}})
\end{equation}
which we can represent by elements as:
\begin{equation}
 E_{kl}=\frac{1}{2}\bigg(\frac{\partial u_k}{\partial X_l}  + \frac{\partial u_l}{\partial X_k} +
 \frac{\partial u_i}{\partial X_k}\frac{\partial u_i}{\partial X_l}\bigg)
\end{equation}
We have cast the Green strain tensor in terms of displacement derivatives.

Hence we can rewrite the Green deformation tensor, ${\bf{C}}$, as follows:
\begin{equation}
 {\bf{C}} = {\bf{1}} + 2{\bf{E}}
\end{equation}
Looking back at \eq{eq:dsds02}, we can rewrite the equation as
\begin{equation}
\label{eq:dsds022}
 \bigg(\frac{ds}{ds_0}\bigg)^2={\bf{e\cdot C \cdot e}}={{\bf{e}} \cdot ({\bf{1}} + 2{\bf{E}})\cdot {\bf{e}}} =
 1 + 2{\bf{e}\cdot \bf{E} \cdot \bf{e}}
\end{equation}
Hence, recalling \eq{eq:longstrain_appendix}, we can also rewrite our longitudinal strain from before in terms of the Green strain tensor.
\begin{equation}
\label{eq:longstrain_appendix2}
\epsilon=\frac{ds}{ds_0}-1= \sqrt{ 1 + 2{\bf{e}}\cdot \bf{E} \cdot \bf{e}} -1
\end{equation}
For the purposes of this paper, this is all that we need to understand in terms of strain theory. 
We will build off these ideas to derive a molecular approach for piezoelectricity.  

\section{Construction of Rotation and Vibration Vectors}
In the Derivation section of this work we mentioned that in order to solve for $\firstDeriv{\uu_{\mathrm{vib}}}{\ff}$ (\eq{eq:finalDxdf}) it is useful to construct a set of vectors 
which span the $3N-6$ (or $3N-5$ for linear molecules) dimensional vibrational space of the molecule or system to exclude rotations and translations from the generated displacement vectors 
in \eq{eq:veq}.  
To understand why this is necessary we must consder the interpretation of the Hessian matrix acting on a displacement vector.  
The Hessian matrix is given by the second derivatives
of the Energy with respect to the nuclear coordinates, $\secondDerivMix{E}{u_i}{u_j}$, or the first derivative of the gradient, $\firstDeriv{\nabla E}{u_j}$. 
When the Hessian matrix acts on a displacement vector it gives the change in force generated by moving the nuclei along the electronic potential energy surface, in the way described by the vector, if the potential was harmonic. 
\begin{equation}
 {\bf{H_{uu}}}\cdot{\bf{u}}=\firstDeriv{\nabla E}{u_j}{du_j}\approx \partial \nabla E 
\end{equation}
At equlibrium, the gradient for the system is at (or approximately at) ${\bf{0}}$, meaning there is no net force acting on the nuclei. 
Therefore a change in the force following displacement of nuclei around the equilibrium is equivalent to the force acting on the nuclei after displacement.  
If we imagine the effect of moving the nuclei in a manner that mimics a rotation or translation, we should not find that any net force now acts on the nuclei (granted that there is no external field or stress acting on the molecule).
This is equivalent to saying that the energy of a molecule or system should only depend on the relative position of its nuclei and not its orientation in space if there
is no external force acting on the system.  
This is obvious for translations but only true for infinitesimal vectors corresponding to rotation.  
As a result, if the equilibrium Hessian acts on these vectors, we expect to 
receive a zero vector as an output.  
This implies that rotational and translational vectors are eigenvectors of the Hessian with zero eigenvalues.  
This is the reason the equilibrium 
Hessian should be singular and why inverting it poses a problem in \eq{eq:xvec}.  

The particular reason the Hessian's singularity poses a problem for neutral molecules or systems is due to rotations.  
A neutral molecule with a dipole has no net translational movement 
in a field, but it will rotate to align the dipole in the direction of the field.  
From \eq{eq:xvec} we see that the change in force on the nuclei caused by the electric field is approximated by the matrix product of the field with the dipole derivative matrix ${\bf{H_{uf}}}$. 
If the resulting vector from this operation were to ``contain`` (which it inevitably does since the dipole interacts with the field) some force generated by rotation, the predicted
displacement vector in the direction of rotation would become infinite, making our calculations no longer useful.  
For this reason we imagine that the molecule does not have any rotational
freedom (as it would likely not in a piezoelectric bulk material), and we instead project \eq{eq:xvec} into the space of the remaining vibrational degrees of freedom.  

In many cases, it is ok to diagonalize the Hessian and omit the 5 or 6 vectors (linear or nonlinear) which correspond to the lowest eigenvalues (usually very close to zero).
The remaining vectors usually correspond to the relative deformation of atoms of the molecule.  
The remaining vectors organized as columns can thus make up the transformation matrix, ${\bf{V}}$, and \eq{eq:finalDxdf} may thus be used to calculate $\firstDeriv{\uu_{\mathrm{vib}}}{\ff}$.
In practice, however, for numerical calculations of the Hessian or for systems with negative eigenvalues (which might be desirable for reasons of representing a smaller subunit
of a bulk material), it is safer to construct rotation and translation vectors and remove them manually from the full basis (which could be the eigenvectors of the Hesssian or any basis
which spans the full $3N$ space. 

As discussed earlier, we can view the vectors the Hessian acts on as displacement vectors for the nuclei.  
In this sense the relative velocities of the nuclei under rotational or translational
motion can be reflected in the displacements.  
To build the translation vectors, it must be true that all of the nuclei must ''move`` or be displaced by the same vector if the energy of the system is not 
to change.  
The molecule also has three dimensions in which it can move.  
To make the vectors orthogonal, we just pick the x, y, and z directions and construct the unit vectors.  
We will assume that the vectors are ordered so that the x,y, and z components of displacement are given for an atom before moving on to the next atom in the vector.  
To illustrate this 
we will assume $u_{ix}$ is the x component of displacement for the $i$th atom in the system, and the full displacement vector for an $N$ atom system has the following form:
\begin{equation}
 {\bf{u}}=(u_{1x}, u_{1y}, u_{1z}, u_{2x}, u_{2y}, u_{2z}, . . ., u_{Nx}, u_{Ny}, u_{Nz})
\end{equation}
Thus the 3 normalized translation vectors for the x, y, and z directions (${\bf{u}}^{tx}$,${\bf{u}}^{ty}$, and ${\bf{u}}^{tz}$ respectively) have the form:
\begin{equation}
 \begin{split}
 {\bf{u}}^{tx}&=(\frac{1}{\sqrt{N}}, 0, 0, \frac{1}{\sqrt{N}}, 0, 0,  . . .,\frac{1}{\sqrt{N}}, 0, 0) \\
 {\bf{u}}^{ty}&=(0, \frac{1}{\sqrt{N}}, 0, 0, \frac{1}{\sqrt{N}}, 0,  . . .,0, \frac{1}{\sqrt{N}}, 0) \\
 {\bf{u}}^{tz}&=(0, 0, \frac{1}{\sqrt{N}}, 0, 0, \frac{1}{\sqrt{N}},  . . .,0, 0, \frac{1}{\sqrt{N}})
 \end{split}
\end{equation}

The rotation vectors are a little trickier to write.  We first need to calculate the geometric center vector for the system and then the position vectors for the nuclei relative to 
the geometric center. 
\begin{equation}
 {\bf{r}}_{gc} = \frac{\sum_i {\bf{x}}_i}{N}
\end{equation}
Here ${\bf{x_i}}$, $N$, and ${\bf{r}}_{gc}$ are the positions of the nuclei, total number of nuclei, and the geometric center vector for the molecule respectively.  
The position vector relative to the geometric center for the $i$th atom is given by
\begin{equation}
\label{eq:geomrel}
 {\bf{r}}_{i}={\bf{x}}_i-{\bf{r}}_{gc}
\end{equation}
It is important to note here that for our purposes we will rotate the molecule about the geometric center and not the center of mass in free rotation.  
We will show later that 
rotating about the geometric center creates vectors orthogonal to the translational vectors. 
For a given angular velocity vector ${\bf{\omega}}$ for a molecule, the velocity vector for an atom in the molecule (${\bf{v}}_i$ is given by 
\begin{equation}
{\bf{v}}_i = {\bf{r}}_{i}\times {\bm{\omega}}
\end{equation}
Thus the displacement vectors we wish to construct have the form
\begin{equation}
\label{eq:rotVec}
 {\bf{u}}^{r\omega}=({\bf{r}}_{1}\times {\bm{\omega}}, {\bf{r}}_{2}\times {\bm{\omega}}, {\bf{r}}_{3}\times {\bm{\omega}}, . . .,{\bf{r}}_{N}\times {\bm{\omega}})
\end{equation}
before normalization.
For a linear or nonlinear molecule, we should be able to construct 2 or 3 rotation vectors respectively that span our space of interest.  
To do so efficiently we will try 
to construct orthongonal vectors. 
We proceed by first realizing that a dot product of two rotation vectors is equivalent to the sum of individual dot products of the displacement vectors
for atoms with the same index.  
There are two different angular velocity vectors (${\bm{\omega}}_j$ and ${\bm{\omega}}_k$) associated with two different rotation vectors.  
\begin{equation}
  {\bf{u}}^{r\omega_j}\cdot {\bf{u}}^{r\omega_k}=\sum_i ({\bf{r}}_{i}\times {\bm{\omega}}_j) \cdot ({\bf{r}}_{i}\times {\bm{\omega}}_k)
\end{equation}
At this point it is helpful to rewrite the cross products as matrix products with the angular velocity vectors. 
The cross product matrix has the form
\begin{equation} 
  \bld{R}_i \equiv 
            \left(
              \begin{array}{ccc}
                0  & -r_{iz}   & r_{iy}     \\
                r_{iz}   & 0   & -r_{ix}     \\   
                -r_{iy} & r_{ix} & 0  \\
              \end{array}
            \right)
\end{equation}
where $r_{ij}$ is the $j$th (x,y, or z) component of the vector ${\bf{r}}_{i}$.

We then replace the cross products with matrix products to obtain
\begin{equation}
   {\bf{u}}^{r\omega_j}\cdot {\bf{u}}^{r\omega_k}=\sum_i ({\bf{R}}_{i} {\bm{\omega}}_j) \cdot ({\bf{R}}_{i}{\bm{\omega}}_k)
\end{equation}
Then recognizing that the dot product is equivalent to a matrix product of a row and column vector, we may take the transform of the matrix product on the left and multiply it by the 
matrix product on the right.
\begin{equation}
 {\bf{u}}^{r\omega_j}\cdot {\bf{u}}^{r\omega_k}=\sum_i ( {\bm{\omega}}^T_j {\bf{R}}^T_{i})({\bf{R}}_{i} {\bm{\omega}}_k)= {\bm{\omega}}^T_j  (\sum_i {\bf{R}}^T_{i}{\bf{R}}_{i}) {\bm{\omega}}_k
\end{equation}
In the last step we have moved the summation over atoms to inside the multiplications with the angular velocity vectors.  
From this we obtain a new matrix.
\begin{equation}
{\bf{S}}= \sum_i {\bf{R}}^T_{i}{\bf{R}}_{i}
\end{equation}
This is a sum over the matrix product shown and is necessarily symmetric.  
Our original goal was to choose angular velocity vectors such that we construct orthogonal rotation vectors.  
Thus we impose 
the requirement that this dot product yields zero.  
\begin{equation}
{\bf{u}}^{r\omega_j}\cdot {\bf{u}}^{r\omega_k}= {\bm{\omega}}^T_j {\bf{S}} {\bm{\omega}}_k=0
\end{equation}
Since our matrix ${\bf{S}}$ is symmetric, it has orthogonal eigenvectors.  Therefore our choice for the set of vectors ${\bm{\omega}}_i$ is obvious.  If we choose the eigenvectors of ${\bf{S}}$ to be 
our angular velocity vectors we have met the conditions.
\begin{equation}
{\bf{u}}^{r\omega_j}\cdot {\bf{u}}^{r\omega_k}= {\bf{\omega}}^T_j {\bf{S}} {\bf{\omega}}_k=\lambda_k{\bf{\omega}}^T_j){\bf{\omega}}_k=0
\end{equation}
Here $lambda_k$ is the eigenvalue for the eigenvector ${\bf{\omega}}_k$.
As we stated before, we chose the geometric center as our rotation point for the system because it also yields rotation vectors orthogonal to the translation vectors.  
This is not hard to show. 
As a matter of fact, we only need the sum of the rotation vectors for each atom per full rotation vector to yield the zero vector--or in equation form:
\begin{equation}
\label{eq:sumtozero}
\sum_i {\bf{u}}^{r\omega_j}_i=\sum_i ({\bf{r}}_{i}\times {\bf{\omega}}_j)={\bf{0}}
\end{equation}
To show that this is the case we substitute \eq{eq:geomrel} into the \eq{eq:sumtozero}.
\begin{widetext}
\begin{equation}
\begin{split}
\label{eq:sumtozerofull}
\sum_i {\bf{u}}^{r\omega_j}_i&=\sum_i ({\bf{r}}_{i}\times {\bf{\omega}}_j)=(\sum_i{\bf{x}}_i-{\bf{r}}_{gc})\times {\bf{\omega}}_j \\
&=(\sum_i({\bf{x}}_i-\frac{\sum_k {\bf{x}}_k}{N}))\times {\bf{\omega}}_j = (\sum_i{\bf{x}}_i-N\frac{\sum_k {\bf{x}}_k}{N})\times{\bf{\omega}}_j \\
&=(\sum_i{\bf{x}}_i-\sum_k {\bf{x}}_k)\times{\bf{\omega}}_j={\bf{0}}\times{\bf{\omega}}_j={\bf{0}}
\end{split}
\end{equation}
\end{widetext}
It might not seem intuitive at first that requiring the sum of the rotation vectors for the indivdual atoms to be the zero vector implies that the full rotation vector is orthogonal to the translation vectors.
Any translation vector (or linear combination of translation vectors) consists of the same vector of displacement per atom (Otherwise the nuclei would not be displaced in the same direction by the same magnitude. 
Therefore we may represent any linear combination of translation vectors as the following:
\begin{equation}
 {\bf{u}}^{t} =a{\bf{u}}^{tx}+b{\bf{u}}^{tx}+c{\bf{u}}^{tx}=(a,b,c, a,b,c, a,b,c, . . .a,b,c)
\end{equation}
The resulting dot product of an arbitrary translation vector with a rotation vector will produce the following expression
\begin{widetext}
\begin{equation}
\begin{split}
 {\bf{u}}^{r\omega_j}\cdot{\bf{u}}^{t}&=\sum_i {\bf{u}}^{r\omega_j}_i \cdot (a,b,c) \\
 &=\sum_i(a u^{r\omega_j}_{ix}+b u^{r\omega_j}_{iy}+c u^{r\omega_j}_{iz})= a\sum_i (u^{r\omega_j}_{ix})+b \sum_i (u^{r\omega_j}_{iy})+c \sum_i(u^{r\omega_j}_{iz}) \\
 &=a*0+b*0+c*0=0
\end{split}
\end{equation}
\end{widetext}
In the second line, we have broken up the individual displacement vectors for the $i$th atom in the displacement vector into the x, y, and z components to break the sum up in the remaining line. 
The last line immediately follows from \eq{sumtozerofull}.  
If the individual displacement vectors for the atoms sum to the zero vector for the rotation displacement vector, then it follows that the 
x, y, and z components sum to zero separately, and the dot product is zero. 
Hence constructing the rotational vectors in this way yields vectors that are internally orthogonal (to one another) and vectors
that are orthogonal to the translation vectors.  

Furthermore we can show that constructing rotation vectors using any center as a rotation point yields a vector that is a linear combination of rotation vectors constructed in the way discussed
above and translation vectors.  
To construct such a displacement vector, we first need to calculate the position vectors of the atoms relative to the center as before.  We will do 
so in such a way that we add a vector ${\bf{s}}$ to the geometric center ${\bf{r}}_{gc}$ to produce the desired arbitrary center of rotation. 
\begin{equation}
\label{eq:geomrels}
 \tilde{{\bf{r}}_{i}}={\bf{x}}_i-({\bf{r}}_{gc}+{\bf{s}})={\bf{x}}_i-{\bf{r}}_{gc}-{\bf{s}}
\end{equation}
We then construct the rotation vector as we have done before--this time using the eigenvectors of the matrix $\tilde{\bf{S}}= \sum_i \tilde{{\bf{R}}}^T_{i}\tilde{{\bf{R}}}_{i}$, 
where the matrix $\tilde{{\bf{R}}}_{i}$ is the matrix form of the cross product of $\tilde{{\bf{r}}_{i}}$ with another vector.  T
he new eigenvectors for $\tilde{\bf{S}}$, $\tilde{\bm{\omega}}_l$, can be
written as a linear combination of the eigenvectors for ${\bf{S}}$, ${\bf{\omega}}_j$
\begin{equation}
 \tilde{\bf{\omega}}_l = \sum_j c_j {\bf{\omega}}_j
\end{equation}
We can now construct a rotation vector, as we have done previously in \eq{eq:rotVec}.
\begin{widetext}
\begin{equation}
 \begin{split}
  {\bf{u}}^{\tilde{r}\tilde{\omega_l}}&=(\tilde{\bf{r}}_{1}\times \tilde{\bm{\omega}}_l, \tilde{\bf{r}}_{2}\times \tilde{\bm{\omega}}_l, \tilde{\bf{r}}_{3}\times \tilde{\bm{\omega}}_l, . . .,\tilde{\bf{r}}_{N}\times \tilde{\bm{\omega}}_l)\\
  &=({\bf{x}}_1-{\bf{r}}_{gc}-{\bf{s}} \times \sum_j c_j {\bm{\omega}}_j, . . .,{\bf{x}}_N-{\bf{r}}_{gc}-{\bf{s}} \times \sum_j c_j {\bm{\omega}}_j) \\
  &=(({\bf{x}}_1-{\bf{r}}_{gc})\times \sum_j c_j {\bm{\omega}}_j-{\bf{s}} \times \sum_j c_j {\bm{\omega}}_j, . . .,({\bf{x}}_N-{\bf{r}}_{gc})\times \sum_j c_j {\bm{\omega}}_j-{\bf{s}} \times \sum_j c_j {\bm{\omega}}_j) \\
  &=(\sum_j c_j {{\bf{r}}_{1}}\times {\bm{\omega}}_j + {\bf{a}}, . . . \sum_j c_j {{\bf{r}}_{N}}\times {\bm{\omega}}_j + {\bf{a}})\\
  &=\sum_j c_j ({\bf{r}}_{1}\times {\bm{\omega}}_j, . . .{{\bf{r}}_{N}}\times {\bm{\omega}}_j)+({\bf{a}}, . . .{\bf{a}})
 \end{split}
\end{equation}
\end{widetext}
Here, ${\bf{a}}=-{\bf{s}} \times \sum_j c_j {\bm{\omega}}_j$, since this vector is constant and does not depend on the atom index.  In the last line, it is clear that the resulting rotational displacement vector
is a linear combination of the rotation vectors about the geometric center plus a vector containing ${\bf{a}}$ for every atom position.  As stated previously, any such displacement vector where
the displacement is constant for every atom can be written as a linear combination of the basal translation vectors.  Therefore we have shown that any displacement vector constructed as a rotation
vector about any point can be written as a linear combination of translation vectors and rotation vectors formed by rotating about the geometric center. Therefore any such set of linearly independent rotation and translation
vectors will necessarily occupy the same space, and after an internal orthogonalizing scheme, can be projected out from the eigenvectors of the Hessian.  The remaining vectors will contain only vibrations.  
This is of course only true for infinitesimal displacements as any vibrational motion of the system will change the rotation and hence vibrational vectors.  For the purposes of this manuscript though we 
are only concerned with the field derivatives of the displacement at zero field, and thus we are not concerned with large system strains.


\end{appendix}


\bibliographystyle{unsrt}

\bibliography{bibliography}

\end{document}